\documentclass[10pt,journal,a4paper]{IEEEtran}
\usepackage{amssymb,amsmath}
\usepackage{cite}
\usepackage{graphicx,subfigure}
\usepackage{psfrag}
\usepackage{url}
\usepackage{soul}
\usepackage[latin1]{inputenc}
\usepackage[absolute,overlay]{textpos}
\usepackage{enumitem}
\usepackage{multirow}
\usepackage{ragged2e}

\usepackage{pgf}
\usepackage[latin1]{inputenc}
\usepackage{bm}
\usepackage{stackengine}

\usepackage{array}
\newcolumntype{L}[1]{>{\raggedright\let\newline\\\arraybackslash\hspace{0pt}}m{#1}}
\newcolumntype{R}[1]{>{\raggedleft\let\newline\\\arraybackslash\hspace{0pt}}m{#1}}

\usepackage{arydshln}
\usepackage{booktabs}
\usepackage{balance}

\usepackage{amsthm}

\usepackage{footnote}
\makesavenoteenv{tabular}
\makesavenoteenv{table}

\begin{document}

\title{Massive Wireless Energy Transfer:\\ Enabling Sustainable IoT Towards 6G Era}

\author{
	\IEEEauthorblockN{	Onel L. A. López, \IEEEmembership{Member, IEEE},
	Hirley Alves, \IEEEmembership{Member, IEEE},
	Richard D. Souza, \IEEEmembership{Senior Member, IEEE},
	Samuel Montejo-Sánchez, \IEEEmembership{Member, IEEE},
	Evelio M. G. Fernández, \IEEEmembership{Member, IEEE},\\
	and Matti Latva-aho, \IEEEmembership{Senior Member, IEEE}
					}						
	\thanks{Onel L. A. L\'opez, Hirley Alves and Matti Latva-aho are with the Centre for Wireless Communications (CWC), University of Oulu, Finland. \{onel.alcarazlopez,hirley.alves,matti.latva-aho\}@oulu.fi}
	\thanks{Richard D. Souza is with Federal University of Santa Catarina (UFSC), Florianópolis, Brazil. \{richard.demo@ufsc.br\}.}
	\thanks{S. Montejo-Sánchez is with Programa Institucional de Fomento a la I+D+i, Universidad Tecnológica Metropolitana, Santiago, Chile. \{smontejo@utem.cl\}.}
	\thanks{E.M.G. Fernández is with Federal University of Paraná (UFPR), Curitiba, Brazil. \{evelio@ufpr.br\}.}
	\thanks{This work is partially supported by Academy of Finland (Aka) (Grants n.319008, n.307492, n.318927 (6Genesis Flagship)), as well as FONDECYT Iniciaci\'on No. 11200659, FONDECYT Regular No. 1201893, and FONDEQUIP EQM180180, in Brazil by the National Council for Scientific and Technological Development (CNPq), and project Print CAPES-UFSC ``Automation 4.0''.}  
}					
					
\maketitle

\begin{abstract}
Recent advances on wireless energy transfer (WET) make it a promising solution for powering future Internet of Things (IoT) devices enabled by the upcoming sixth generation (6G) era. The main architectures, challenges and techniques for efficient and scalable wireless powering are overviewed in this paper.
Candidates enablers such as energy beamforming (EB), distributed antenna systems (DAS), advances on devices' hardware and programmable medium, new spectrum opportunities, resource scheduling and distributed ledger technology are outlined.
Special emphasis is placed on discussing the suitability of channel state information (CSI)-limited/free strategies when powering simultaneously a massive number of devices. 
The benefits from combining  DAS and EB, and from using average CSI whenever available, are numerically illustrated. 
The pros and cons of the state-of-the-art CSI-free  WET techniques in ultra-low power setups are thoroughly revised, and some possible future enhancements are outlined.
Finally, key research directions towards realizing 
	WET-enabled massive IoT networks in the 6G era are identified and discussed in detail.
%
%
\end{abstract}
\section{Introduction}\label{int}
%
%

The sixth generation (6G) of wireless systems targets a data-driven sustainable society, enabled by near-instant, secure, unlimited and green connectivity \cite{ZhangXiao.2019,MahmoodAlves.2020,Mahmood.2020}. Stringent performance requirements in terms of security and trust, throughput, sensing capabilities, dependability, scalability and energy efficiency, as illustrated in Fig.~\ref{figInt}, have been set by industry and academy to fulfill such a vision.
Specifically, the ultimate vision in terms of energy efficiency is that of 
a green society assisted by 6G networks, specially by zero-energy/cost/emission
Internet of Things (IoT) deployments \cite{Mahmood.2020,Huang.2019}.  
However, this is still
 a major concern due to the lack of mature solutions for powering and keeping uninterrupted operation of the massive number of devices. 

Technological advances on artificial intelligence (AI)\!\! /machine learning (ML), molecular, backscatter and visible light communications, fog/edge computing, and metamaterials/metasurfaces will certainly facilitate sustainability \cite{Mahmood.2020,Huang.2019,Akyildiz.2020}. In addition to these,
research community and industry are considering energy harvesting (EH) techniques an attractive solution to externally recharge batteries or avoid replacement \cite{Bi.2015,Lu.2015,Tran.2017,Divakaran.2019}, which may be not only costly but also impossible in hazardous environments, building structures or the human body. 
Therefore, EH is foreseen as a key component of future IoT networks since it allows \textit{i})
%
	wireless charging, which significantly simplifies the servicing and maintenance of IoT devices, while increasing their durability thanks to contact-free feature; and
	\textit{ii}) enhanced energy efficiency and network-wide reduction of emissions footprint. Notice that the battery recharging and waste processing is	already a critical problem for which EH is an attractive clean solution \cite{Tran.2017,Portilla.2019}. 
%
\subsection{EH Technologies}\label{Sources}
\begin{figure*}
	\centering
	\includegraphics[width=1.9\columnwidth]{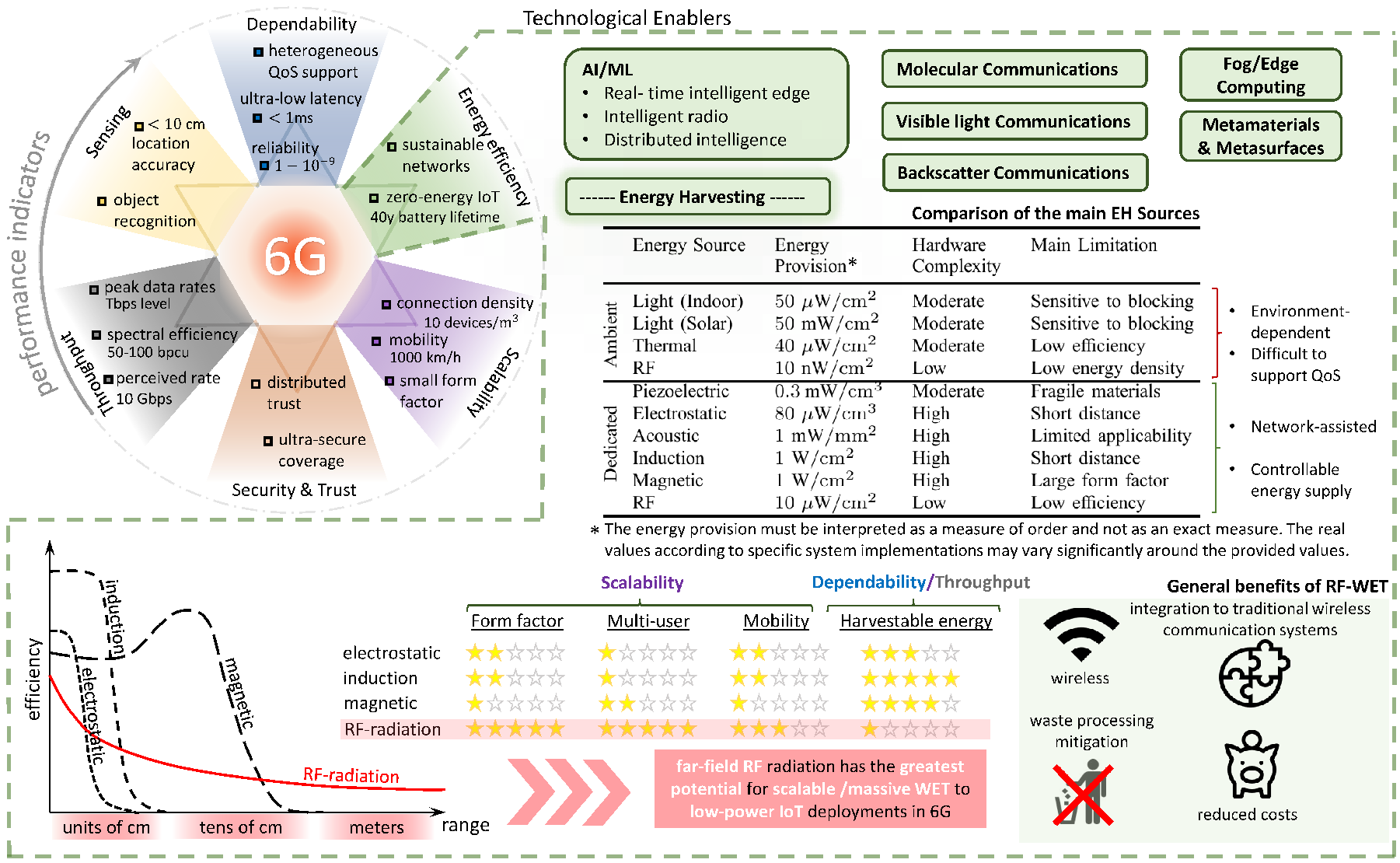}
	\caption{Radio frequency (RF)-WET as key enabler of energy-efficient and scalable 6G networks. Specifically, the figure summarizes 6G performance requirements \cite{ZhangXiao.2019,MahmoodAlves.2020,Mahmood.2020}, technological enablers of energy efficiency in 6G \cite{Mahmood.2020,Huang.2019,Akyildiz.2020}, EH techniques \cite{Prauzek.2018,Chen.2019,Awal.2016}, characteristics and advantages of RF-WET \cite{Bi.2015,Cheah.2019}. Regarding the latter, a comparison (in terms of coverage, form factor, harvestable energy, and multi-user and mobility native support) between RF-WET and main competitors (electrostatic or capacitive coupling, inductive coupling and magnetic coupling) is depicted \cite{Bi.2015,Cheah.2019}.  All in all, RF-WET constitutes the most prominent technology for massively and wirelessly powering low-energy IoT deployments.}
	\label{figInt}
\end{figure*}

EH technologies can be classified into the following two categories \cite{Bi.2015}:
\begin{itemize}
	\item ambient EH, which relies on  energy resources that are readily available in the environment and that can be sensed by EH receivers;
	\item dedicated EH, which are characterized by on-purpose energy transmissions from dedicated  energy sources to EH devices.
\end{itemize}
Different from dedicated EH setups, ambient EH does not require additional resource/power consumption from the surrounded (sometimes newly-deployed) energy network infrastructure. However,  temporal/geographical/environmental circumstances may limit their service guarantee making them inappropriate (at least as standalone) for many use cases with quality of service (QoS) requirements. The main energy sources within the above categories and their associated characteristics are summarized in Fig.~\ref{figInt}.
The ambient EH methods based on light intensity, thermal energy or even wind, are either highly sensitive to blocking or perform with low conversion efficiency \cite{Prauzek.2018}. But maybe more importantly, they demand an add-on EH material \& circuit, which in practice limits the form factor reduction to the desired levels  for many use cases. 
The same strong limitation is characteristic of the induction, magnetic resonance coupling and many piezoelectric-based dedicated EH methods. The induction method is based on the inductive coupling effect of non-radiative electromagnetic fields, including the inductive and capacitive mechanisms,
and is subject to coupling misalignment impairments that limit the range and scalability \cite{Chen.2019}.
The magnetic resonance coupling 
exploits the fact that 
two objects resonating at the same frequency tend to couple with each other most efficiently. In fact, by carefully tuning the transmitter and receiver circuits, magnetic resonant coupling is able to achieve higher power transfer efficiency over longer distances than inductive coupling  \cite{Chen.2019}.
Meanwhile, piezoelectric-based EH relies on the energy coming from a mechanical strain captured by a, usually fragile, piezoelectric material layer on top of the wireless device \cite{Chen.2019}.
In general, electrostatic and acoustic methods overcome the devices' size limitation, but they are either limited to very short distance operations or to very specific applications. Specifically, in the electrostatic method, a mechanical motion or vibration is used to change
the distance between two electrodes of a capacitor against an electric field, thus, transforming the vibration or motion into electricity due to the capacitance change \cite{Chen.2019}. 
Meanwhile, 
acoustic energy transfer, which is usually in the range of ultra-sound, is
quite efficient but mostly for transferring the energy over non-aerial media such as water, tissue
or metals, and it is more appropriate for medical applications \cite{Awal.2016}.
It is worth pointing to  laser power beaming as another potential EH technology, which uses highly concentrated
laser light aiming at the EH receiver to achieve efficient power delivery over long distances \cite{Zhang.2019}.
However, this technology may be just suitable for powering  complex devices with high power consumption demands, e.g., smart phones, while it requires accurate pointing towards the receiver.
\begin{table*}[!t]
	\centering
	\caption{Brief Summary on Existing Surveys and Overviews of RF-EH/WET (2015-2020)}
	\begin{tabular}{p{0.4cm}p{0.6cm}p{3.1cm}p{12.3cm}}
		\toprule
		 \textbf{Year} & \textbf{Ref.} & \textbf{Focused issues} & \textbf{Main contents}  \\
		\midrule
		2015 & \cite{Bi.2015}   &  wireless powered communications &  $\vartriangleright$RF-enabled WET technologies \& their applications to wireless communications, --network models, --signal processing methods to enhance WET efficiency, --WET \& WIT integration trade-offs, --design challenges, solutions,
		and opportunities ahead   \\ 
		& \cite{Lu.2015} &  wireless networks with RF EH & $\vartriangleright$architecture of wireless powered networks, --RF EH techniques and propagation models, --circuit design, --energy beamforming, resource allocation and scheduling, RF-powered cognitive radio networks, --routing and medium access control (MAC) protocols, --research directions\\
		& \cite{Kosunalp.2015} & MAC protocols for EH sensor networks  & $\vartriangleright$sources for ambient EH, --network architecture, --power management, --MAC protocols, --research directions \\ \hdashline
		2016 & \cite{Bi.2016} & wireless-powered communication networks (WPCNs) & $\vartriangleright$WPCN basic models, --key enablers (energy beamforming, joint communication and network scehduling, cooperation), --research directions and practical challenges \\	
		& \cite{Ghazanfari.2016} & ambient RF EH in ultra-dense networks & $\vartriangleright$feasibility and challenges, --energy arrival models, --network operation and energy management,  --design considerations, --performance metrics, --numerical examples\\
    	& \cite{Soyata.2016} & RF EH for embedded systems & $\vartriangleright$EH sources, --RF EH applications, fundamentals, circuits and performance, --power management, --WET \& WIT integration principles and trade-offs \\ \hdashline
		2017 & \cite{Ramezani.2017}  & WPCNs & $\vartriangleright$network models (large scale deployments, cognitive and relying networks), --enhancing techniques (full-duplex, multi-antenna WET \& WIT), --wireless powered IoT, --future research possibilities  \\
		& \cite{Hu.2017} &  WET \& WIT integration in  body area networks & $\vartriangleright$WET in body area networks, --single point-to-point WET \& WIT integration, --network models, --design challenges and opportunities\\
		& \cite{Mekid.2017} & feasibility studies of ambient RF EH & $\vartriangleright$trials of EH systems in industrial market, --loss circuit analysis, --spectral analysis, --experimental test \\
		& \cite{Niyato.2017} & WPCNs & $\vartriangleright$wireless EH technologies, --network architecture and standard
		development, --implementation perspectives, --backscatter communications with EH, --energy management, --EH aware transceiver and algorithm design, --research directions\\ \hdashline
		2018 & \cite{Alsaba.2018}  & beamforming in wireless powered communications & $\vartriangleright$basic concepts \& architecture of wireless powered networks, --multi-antenna EH-enabled communications, --beamforming transmission schemes, --physical layer security, --advances, open issues, challenges, and future research directions \\ 
		& \cite{Tran.2018} & wireless powered networks & $\vartriangleright$green WET, --EH models, --future green networks with EH, --green radio communications (full-duplex, millimeter wave (mmWave), wireless sensor networks, cooperation), --heterogeneous networks, --research directions \\
		& \cite{Perera.2018} & wireless powered communications & $\vartriangleright$RF EH principles (circuit design, applications, challenges), --WET variants, --simultaneous wireless information and power transfer (SWIPT) architectures, --interference exploitation in SWIPT, --emerging SWIPT schemes, --research directions  \\ 
		& \cite{Hu.2018} & WET \& WIT integration & $\vartriangleright$ambient EH, --near/far field WET,  --architectures for WET \& WIT integration, --information theory principles, --hardware implementation, --transceiver design, --energy shortage, --resource allocation, --MAC design, --connecting WET \& WIT users, --challenges and research directions \\
		& \cite{Huynh.2018} & ambient backscatter communications & $\vartriangleright$principles and fundamentals, --applications and challenges, --bistatic backscatter communications systems, --emerging backscatter systems, --research directions \\\hdashline
		2019 & \cite{Divakaran.2019} & RF EH systems  & $\vartriangleright$antenna design, --rectifier and matching network, --trends in rectennas, --efficient design methods \\
		& \cite{Baroudi.2019} & energy management in ambient RF EH networks & $\vartriangleright$ambient energy sources, --taxonomy of energy management, --soft power management, --research challenges\\
		& \cite{Liang.2019} & SWIPT in 5G networks & $\vartriangleright$SWIPT systems, --Non-orthogonal multiple access (NOMA) and cooperation, --SWIPT-enabled mmWave 5G network, --research challenges  \\ \hdashline
		2020  	& \cite{LiuDai.2020} & unmanned aerial vehicle (UAV)-enabled wireless powered IoT & $\vartriangleright$system working flow and design challenges, --enabling technologies (adaptive energy beamforming and IoT EH circuit design, resource allocation and optimization, UAV trajectory optimization), --future research directions \\    	
    	 & \cite{Hu.2020} & EH in
    	 wireless networks interoperating with smart grids & $\vartriangleright$classical models of EH technologies, --operation and optimization, --WET \& WIT, --redistribution of redundant energy harvested within cellular networks, --energy planning under dynamic pricing in smart grids, --two-way energy trading between cellular networks
    	 and smart grids, --application of optimization tools, --future potential applications, --research directions \\
    	  & \cite{Memon.2020} & ambient backscatter communications & $\vartriangleright$energy sources, --frequency band conversion, --channel estimation and transmission	distance, --power transfer management, --scheduling and resource allocation, --system performance, --research directions \\
    	  & \cite{Duan.2020} & ambient backscatter communications for ultra-low-power IoT  & $\vartriangleright$characteristics and applications, --challenges and solutions (direct path interference, receiver design, coherent and non-coherent receivers, intelligent receivers), --research trends   \\
    	  & \cite{Nawaz.2020} & non-coherent \& backscatter communications & $\vartriangleright$requirements of beyond 5G networks, --non-coherent communications, --principles, design issues,  and revision of
    	  backscatter communications, --ultra massive connectivity in the 6G era, --open challenges and potential research topics \\
		\bottomrule
	\end{tabular}\label{tableS}
\end{table*} 
\subsection{Scope and Contributions of this Work}\label{cont}
In contrast to the above discussed EH methods, RF-based EH inherently allows:
\begin{itemize}
	\item small-form factor implementation. Devices' dimensions have been determined by the size of traditional batteries, while RF-EH batteries (if needed) are smaller \cite{Tran.2017}. Additionally, 
	the same RF circuitry for wireless communications can be re-utilized totally or partially for RF EH;
	\item native multi-user support  since the same RF signals can be harvested simultaneously by several devices.
\end{itemize}
The above key features, when combined, make RF EH a strong candidate, much more suitable than the EH technologies based on other energy sources,  for powering many low-power IoT use cases. 
When the number of devices increases, RF-EH technologies become even more appealing. Fig.~\ref{figInt} also illustrates a comparison between radiative RF wireless energy transfer (WET) and the main non-radiative competitors, and highlights the native advantages of RF-WET for powering massive low-energy IoT deployments.

There is a wide range of recent literature surveying and/or overviewing RF-EH related topics. In this regard, Table~\ref{tableS} summarizes the focus and content of the main surveys and overviews of RF-EH in the last six years. 
Readers may refer to \cite{Kosunalp.2015,Ghazanfari.2016,Mekid.2017,Baroudi.2019,Divakaran.2019} for an overview of ambient RF EH, to \cite{Niyato.2017,Huynh.2018,Duan.2020,Memon.2020,Nawaz.2020} for revisions of future ambient backscatter communications, and to \cite{Bi.2015,Lu.2015,Kosunalp.2015,Bi.2016,Soyata.2016,Ramezani.2017,Hu.2017,Niyato.2017,Alsaba.2018,Tran.2018,Perera.2018,Hu.2018,Liang.2019,LiuDai.2020,Hu.2020} to explore RF-WET and wireless information transmission (WIT) integration issues.
Meanwhile, the focus in this work is on dedicated RF WET, hereinafter referred just as WET, which is the most
critical building block in practical RF-EH communication systems. This is because enabling devices to harvest sufficient amount of energy for operation and communication is extremely challenging and causes the system performance bottleneck in practice.
 Also, different from all above works, the emphasis here is on the system characteristics, enablers and challenges for powering massive IoT deployments in the 6G era.
Extensive discussions are carried out on the emerging need of efficient channel state information (CSI)-limited/free WET mechanisms\footnote{Decades ago, before the emergence of adaptive modulation and	coding/pre-coding schemes, non-coherent (CSI-free) wireless transmissions were commonplace. Nowadays, however, the stringent performance requirements of beyond 5G and 6G systems challenge the design of practical CSI-limited schemes \cite{Lopez.2020_mag}. Readers are encouraged to refer to \cite{Nawaz.2020,Xu.2019} for a thorough revision and overview of non-coherent communications, recent advances, research trends and application to future backscatter communications scenarios.
}.
	%
%
\begin{table}[!t]
	\centering
	\caption{List of Acronyms}
	\begin{tabular}{ll}
		\toprule
		\textbf{Acronym} & \textbf{Definition}  \\
		\midrule
		AG  & Average gain  \\ 
		AI  & Artificial intelligence  \\ 
		APS & Antenna phase shifting\\
		BS  & Base station \\
		CMOS & Complementary metal-oxide-semiconductor \\
		CSI & Channel state information\\
		DAS & Distributed antenna systems   \\
		DG & Diversity gain \\
		DLT & Distributed ledger technology \\
		EB & Energy beamforming\\
		EMW & Energy modulation/waveform\\
		IoT & Internet of Things\\
		IRS & Intelligent reflecting surfaces\\
		ISM & Industrial, scientific and medical\\
		LIS & Large intelligent surface\\
		LOS & Line of sight\\
		MAC & Medium access control\\
		MIMO & Multiple-input multiple-output\\
		ML  & Machine leaning  \\
		mMIMO & Massive MIMO \\
		mmWave & Millimeter wave \\
		mWET & Massive WET \\
		NOMA & Non-orthogonal multiple access\\
		oDAS & Optimized DAS\\
		PB & Power beacon\\
		PIN &  Positive-intrinsic-negative \\
		PMU & Power management unit \\		
		QoS & Quality of service \\
		RF & Radio frequency\\
		SDP & Semi-definite program\\
		SIC & Successive interference cancellation\\ 
		SINR &  Signal-to-noise-plus-interference ratio\\
		SNR & signal-to-noise ratio \\		
		SWIPT & Simultaneous wireless information and power transfer\\
		UAPB & unmanned aerial PB\\
		UAV & unmanned aerial vehicle\\
		ULA & Uniform linear array\\
		WET & Wireless energy transfer\\
		WIT & Wireless information transfer\\
		WPCN & Wireless-powered communication network\\
		WSN & Wireless sensor networks\\
		WUR & Wake-up radio\\
		\bottomrule
	\end{tabular}\label{tableA}
\end{table} 
\begin{figure}
	\centering
	\includegraphics[width=0.85\columnwidth]{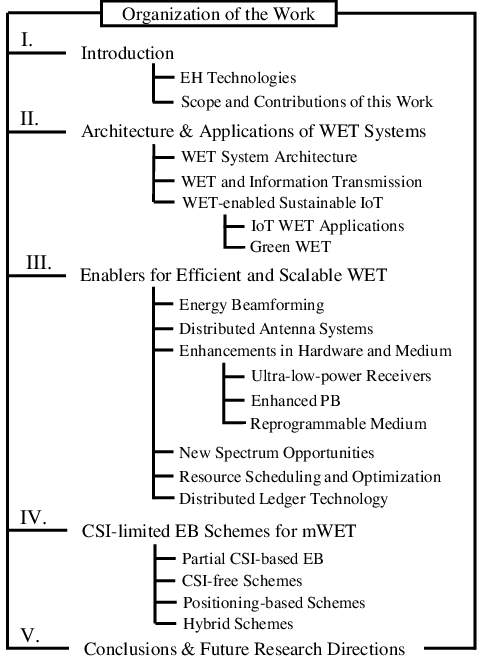}
	\caption{Organization of this work.}
	\label{figTax}
\end{figure}
Specifically, our main contributions are two-fold:
\begin{itemize}
	\item the main features, potentials and 
		challenges of WET systems
	toward powering massive IoT deployments in the 6G era, i.e., massive WET (mWET), are overviewed and highlighted. Our discussions focus specifically on architecture and use cases of WET-enabled networks, and candidate enablers for efficient and scalable WET such as  energy beamforming (EB), novel distributed antenna systems (DAS), advances on devices' hardware and programmable medium, new spectrum opportunities, resource scheduling and distributed ledger technology (DLT). It is shown numerically\footnote{Extensive Monte Carlo simulations in MatLab software were carried out to illustrate the numerical performance trends discussed throughout the paper. The code scripts are publicly available at \url{https://github.com/onel2428/WEToverview}.} that by intelligently deploying DAS, the system performance boosts up compared to that under EB from collocated antennas or un-optimized DAS with the same  number of transmit antennas and transmit power. However, by combining DAS and EB, the deployment costs can be significantly reduced  without significantly compromising the system performance;%
	\item CSI acquisition is discussed and identified as a strong limitation towards mWET. Note that low-cost accurate CSI acquisition is already challenging in WET-enabled small-scale networks \cite{Ramezani.2017}, while the problem exponentially scales up with the network size \cite{Lopez.2020_mag}. 
	In that regard,  novel solutions that do not rely on instantaneous CSI availability are overviewed. Specifically, it is shown that an EB based on average CSI can attain near optimum performance in WET setups. Meanwhile, the advantages and limitations of state-of-the-art CSI-free techniques, along with possible future enhancements, are also discussed and illustrated.
\end{itemize}

Table~\ref{tableA} lists the acronyms used throughout the paper in alphabetical order. 
The organization of the paper is
depicted in Fig.~\ref{figTax}. Specifically, Section~\ref{wet} presents the general architecture of WET systems along with the main IoT-related use cases. Different enablers for efficient and scalable WET-enabled networks are discussed in Section~\ref{enablers}, while Section~\ref{scenarios} overviews specific CSI-free solutions for powering massive low-power IoT deployments. Finally, Section~\ref{conclusions} concludes the paper.  
\section{Architecture \& Applications of WET Systems}\label{wet}
WET is currently being considered, analyzed and tested 
as a stand-alone incipient technology, though its wide integration to the main wireless systems seems unavoidable. 
Cellular industry is continuously expanding its service provision horizons: 1G--2G systems were mainly focused on voice communications; 3G--4G expanded to broadband connectivity provision; while 
5G is not only aiming at ultra-broadband connectivity but also targets industrial and IoT use cases, whose requirements diverge from those of traditional human-type communications. Moreover, 6G is expected to arrive by 2030 with an even stronger service portfolio for large, medium and small-scale industries, thus, shifting to a dramatic verticalization of the service provision \cite{Mahmood.2020}. Wireless energy supply will be undoubtedly explored since it is a lucrative and expanding market
\footnote{See https://www.powercastco.com, https://www.transferfi.com and https://www.ossia.com.}.
\subsection{WET System Architecture}
\begin{figure}
	\includegraphics[width=0.9\columnwidth]{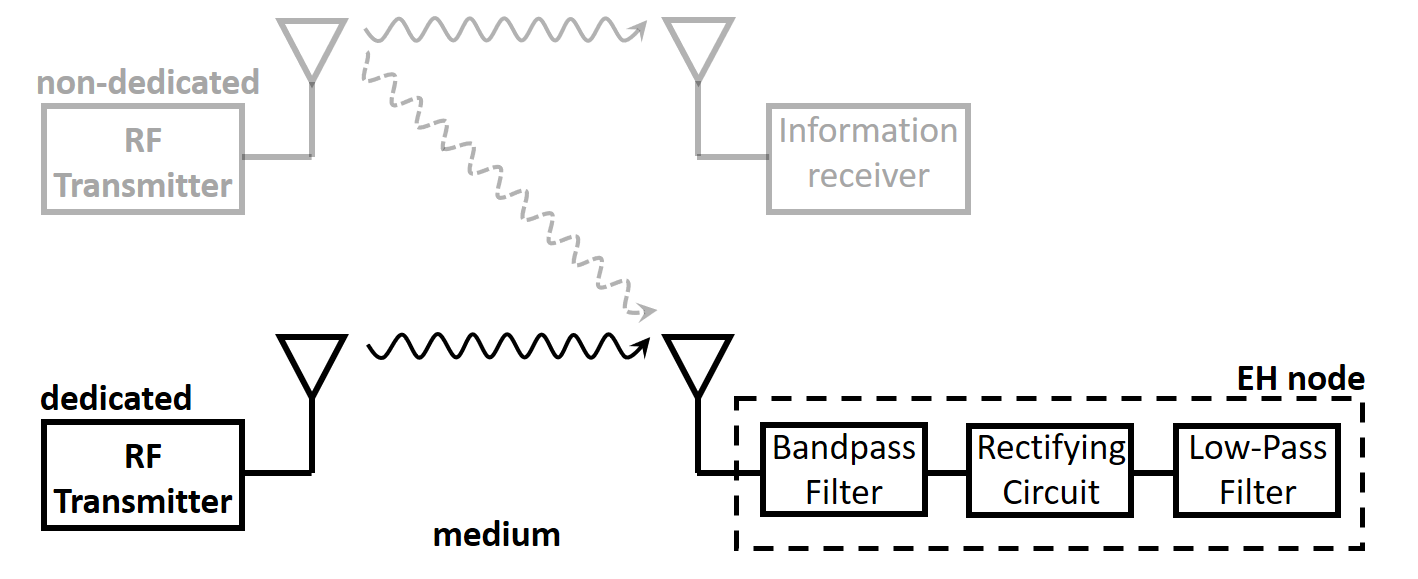}
	\caption{WET system. RF-EH module structure.}
	\label{fig2}
\end{figure}

The basic structure of a WET system is illustrated in Fig.~\ref{fig2}. RF EH devices are equipped with an energy conversion circuit consisting of \cite{Valenta.2014}:
\begin{itemize}
	\item receive antenna(s), which can be designed to work on either single frequency or multiple frequency bands such that the EH node can harvest from single or multiple sources simultaneously. Nevertheless, the RF energy harvester typically
	operates over a range of frequencies since energy density of RF signals is diverse in frequency;
	\item a combination matching network/bandpass 
	filter, which consists of a resonator circuit operating
	at the designed frequency to maximize the power transfer
	between the antenna and the rectifier. It ensures that the harmonics generated by the rectifying element are not re-radiated to the environment and that the efficiency of the impedance matching is high at the designed frequency;
	\item a rectifying circuit, which is based on diodes and capacitors. Generally, higher conversion efficiency can be achieved by diodes with lower built-in voltage. The capacitors ensure to deliver power 	smoothly to the load. 
	\item a low-pass filter, which removes the fundamental and harmonic frequencies from the output and sets the output impedance.  
\end{itemize}
Note that when RF energy is 	unavailable, the capacitor(s) in both the rectifying circuit and low-pass filter can also serve as a reserve for a short duration \cite{Lu.2015}.


The semiconductor-based rectifier is quite common because of its low cost and small-form factor. Specifically, complementary metal-oxide-semiconductor
(CMOS) technology with diode-connected transistors are mostly adopted to significantly increase the EH efficiency at lower powers because of lower parasitic values and customizable rectifiers. This technology is particularly suitable for very low-power low-cost applications such as RFID since the entire device can be incorporated on a single integrated circuit. 
Meanwhile, rectifying antenna (rectenna)-based designs are preferable when larger power densities are available.
Readers may refer to \cite{Valenta.2014,Divakaran.2019} for a detailed overview of the main rectenna designs and EH circuit topologies.
%
\subsection{WET and Information Transmission}\label{WETWIT}
The IoT paradigm includes WIT at heart, hence WET appears naturally combined with WIT either in a WPCN or SWIPT setup. In a WPCN, an RF transmitter powers the EH device(s) via WET in a first phase.
In a second phase, the energy harvested by the device(s) is	
used completely or partially 
for WIT to an information receiver, which may, or may not, be the same node that initiated the WET process. 
Meanwhile, WET and WIT occur  in the same link direction in a SWIPT setup. In this case, the RF source transmits both energy and information signals to either separate or co-located EH and information receivers.
Note that WET, WPCN and SWIPT are canonical models/modes that may appear combined, or even intermittently, in the same network system. For instance, a certain RF transmitter may be powering nearby EH IoT devices that are performing sensing tasks (system in WET mode). At intervals, the RF transmitter requests  information updates from its associated EH devices via WIT (system in SWIPT mode), which then send back the response data messages (system in WPCN mode).

\begin{figure}
	\includegraphics[width=0.95\columnwidth]{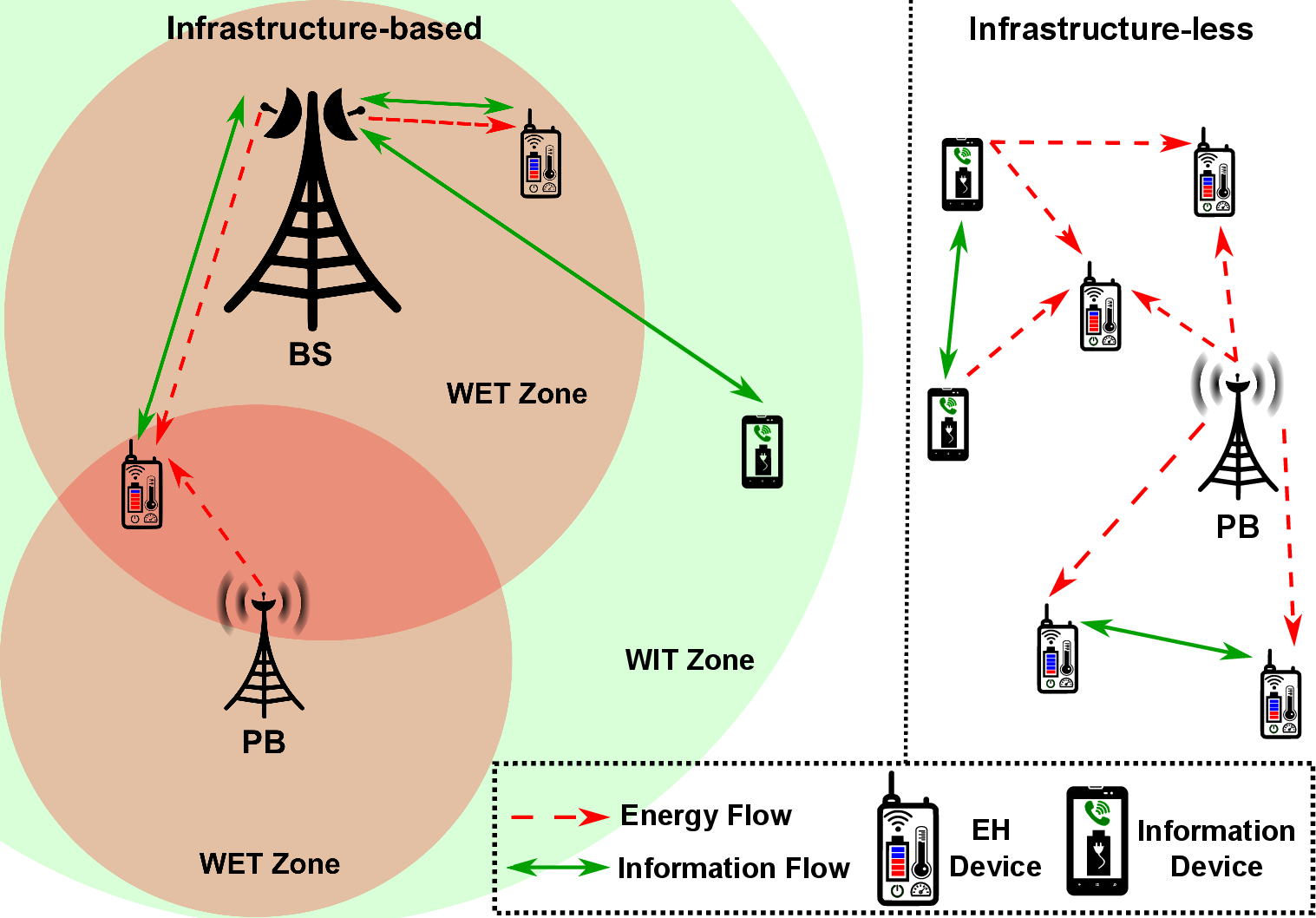}
	\caption{General architectures of WET-enabled networks (adaptation from \cite[Fig.~2]{Lu.2015}).}
	\label{fig3}
\end{figure}

Typical infrastructure-based/less architectures for a WET-enabled network are illustrated in
Fig.~\ref{fig3}.
In an infrastructure-based architecture, there are three major components \cite{Lu.2015}: 
\begin{itemize}
	\item the information gateways, which are generally known as base stations (BSs), wireless routers and relays;
	\item the RF energy transmitters, the so-called power beacons (PBs)\footnote{RF energy transmitter may be also ambient RF sources, e.g., TV towers, as discussed in Section~\ref{Sources}, but it is not the focus here.};
	\item the network nodes/devices, which are the user equipments with/without RF EH capabilities and which  may communicate with the information gateways.
\end{itemize}
Typically, the information gateways and RF energy transmitters have continuous and fixed power supply but they can also/conversely rely on EH from ambient energy sources to promote sustainability (see Section~\ref{green}), or even from other dedicated energy transfer processes, e.g., laser power beaming from higher-hierarchy nodes, while the network nodes harvest energy from RF sources to support their operation.

The EH and WIT zones are illustrated in Fig.~\ref{fig3}. The devices in the EH zone of the information gateway are able to harvest RF energy, while the devices in the WIT zone can successfully decode information transmitted from the gateway; in both cases with certain QoS guarantees. Note that WIT zones are greater than EH zones since the circuitry for
energy and information transmissions operate with very different sensitivity levels \cite{Bi.2015}. While typical information receivers can operate with sensitivities ranging from $-130$ dBm to $-60$ dBm receive signal power; an EH device needs usually more than $-30$ dBm. The reason is that for information decoding the metric of interest is based on a ratio, e.g., the signal-to-noise ratio (SNR) or  signal-to-noise-plus-interference ratio (SINR), and what it matters is how stronger/weaker is the signal power compared to the noise (+interference) level; 
while the nominal received power is what matters for EH purposes. These constitute also the main reasons motivating the deployment of PBs in the first place, i.e., to increase WET coverage to the desired regions. In practice, WET zone dimensions depend on the network scenario and channel propagation conditions, transmit and receive hardware capabilities, and transmission strategies. Finally, note that PBs and BSs can also incorporate WIT and WET functionalities, respectively, i.e., the so called hybrid PBs and hybrid BSs.

In the infrastructure-less architecture, on the other hand, there is no infra-structure-based information gateways, and WIT occurs peer-to-peer. Some peer-to-peer WIT transmissions may be important sources of RF energy for nearby EH devices, which may also be served by ambient and dedicated RF energy transmissions.
\begin{figure}[t!]
	\centering
	\includegraphics[width=0.48\textwidth]{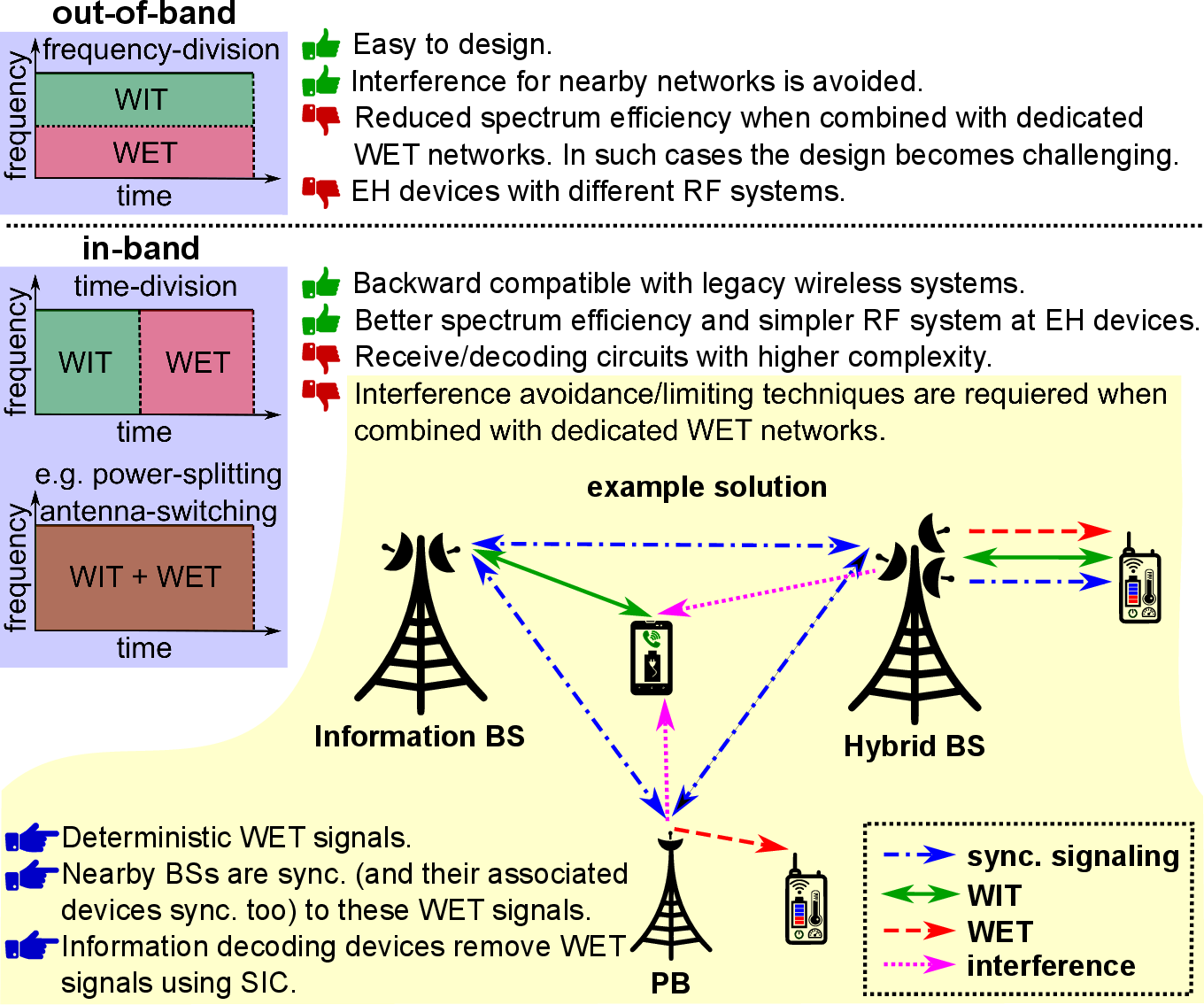}
	\caption{Out-of-band and in-band WET schemes. Advantages and disadvantages. Example of in-band WET using successive interference cancellation (SIC) for removing the interference caused by WET signals to the information decoding devices.}	
	\label{Fig2}
\end{figure}

%
%

Note that 
WET is a fundamental and sensitive building block in these networks since its duration could be larger than WIT in order to harvest usable amounts of energy \cite{Lopez.2017,Lopez.2018_2}. Some cases require perennial WET operation while WIT happens sporadically, e.g., due to event-driven traffic. Dedicated RF transmitters or PBs may be mandatory in many cases.
Additionally, due to the heterogeneity of EH and information decoding circuits,
different 
antenna and RF systems are usually required \cite{Bi.2016}. Moreover, energy and information transmission can be performed either in an out-of-band or in-band
manner.
Although out-of-band and in-band WET schemes are plausible, see Fig.~\ref{Fig2}, the latter is preferable in terms of spectral efficiency when EH and non-EH devices coexist. A solution for mitigating the interference from WET signals to nearby/coexisting information networks is also  illustrated at the bottom of Fig.~\ref{Fig2}. The key idea lies in using deterministic (known by the network) WET signals, which the information decoding devices can mitigate using SIC. Synchronization/coordination between the PBs and BSs may be needed in order to maintain  the associated devices with updated information of the network WET signals.
Finally, in-band schemes allow self-energy recycling in multi-antenna nodes, which benefit them with a secondary energy source when the receive antennas are used for EH \cite{Maso.2015,Lopez.2020_2}. 
%
%
\subsection{WET-enabled Sustainable IoT}\label{iotApp}
6G systems aim at supporting massive connectivity up to the order of 10 devices$/\mathrm{m}^3$ \cite{Mahmood.2020}, most of which will be low-cost low-power IoT devices. WET is an attractive technology for wirelessly and sustainably powering many of such massive IoT networks. Note that 
WET favors network sustainability
by i) simplifying servicing and maintenance, ii) increasing IoT devices' durability, iii) supporting network-wide reduction of emissions footprint by mitigating the battery waste processing problem, and iv) allowing miniaturized hardware implementations and supporting  multi-user operation, thus, favoring deployment scalability.
\subsubsection{IoT WET Applications}
Owing to the above inherent advantages of RF EH and combined with the dwindle in devices' operating power and 
the 
developments in multiple-input multiple-output (MIMO) technology, an increasing number of IoT WET applications is expected towards the 6G era. 
Note that optimizations for one application may be detrimental to another, sacrificing input power range for 
conversion efficiency, size for durability, or complexity for cost in ways suited only to a specific task \cite{Soyata.2016}. Thus, application requirements must be fully considered to ensure circuit trade-offs are given proper weight in the 
final product. 
Some specific applications for WET in the IoT era are illustrated in Fig.~\ref{Fig1}, and include:

\begin{figure}[t!]
	\centering
	\includegraphics[width=0.48\textwidth]{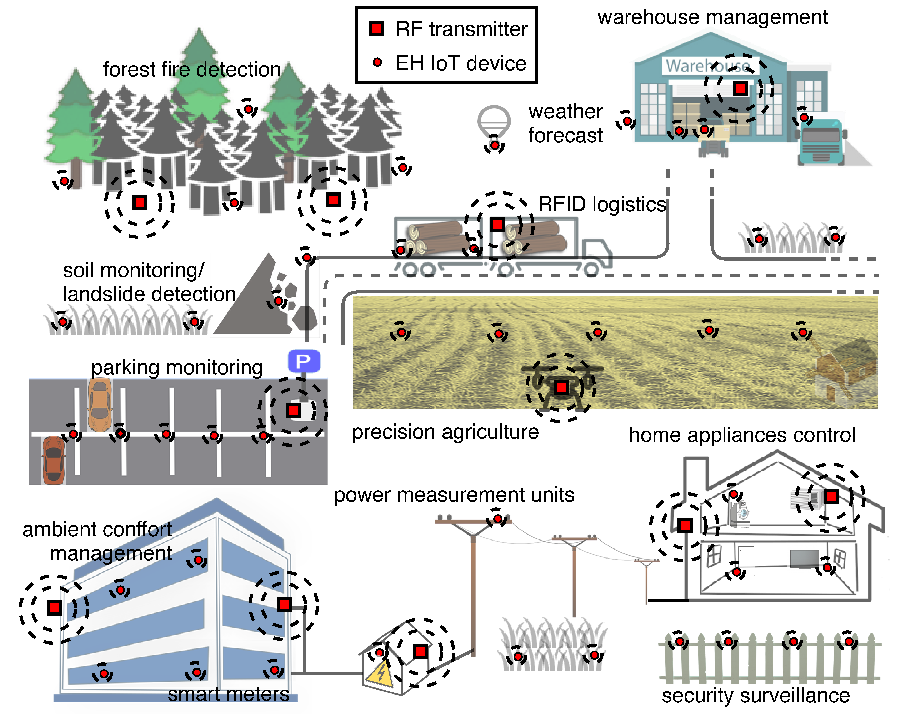}
	\caption{Selected WET-powered IoT use cases, e.g., 
		monitoring/control systems in smart homes; transportation, healthcare, 
		smart grid, weather forecast,
		emergency detection/response, soil monitoring and} precision agriculture. 
	\label{Fig1}
\end{figure}
\begin{itemize}
	\item RFID, which is a key technology  in identification, tracking, and inventory management distributed applications.  
	By incorporating RF EH to the RFID tags, i.e., active or hybrid RFID tags, the  lifetime, functionality and/or operation range is considerably extended  compared to traditional designs \cite{Lu.2015};
\item live labels. The labels of future products (in shops and markets) can be designed to provide information such as pricing, flash special offers, bar codes and  freshness index. Even simple user feedback may be sensed and sent to the shop/manufacturer. Tags with light-based EH and optical access points in ceilings are nowadays been considered \cite{Katz.2019}, however, WET or hybrid architectures may be preferred to avoid blocking;	
\item wireless sensor networks (WSNs), which cover applications from smart house, healthcare to industry and military. 
	Applying RF energy to recharge or avoid the need of batteries is one promising approach to enhance the lifespan of WSNs\footnote{WSN deployments often promote sustainability by their own, e.g., when motion-tracking sensors intelligently turn on or off the lights given that someone entered, or everyone left, a room, thus saving important energy resources.}. For instance, authors in \cite{Ruisi.2016} proposed WET for supporting the  structural monitoring of buildings. Specifically, far-field RF EH sensors were developed to detect humidity, temperature and light inside a building. Results showed that at a distance of 1 m from a 3 W source, the sensor node received 3.14, 2.88, 1.53, 0.7 mW power through air, wood, 2 inches of brick and 2 inches of steel, respectively which is sufficient for powering many state-of-the-art sensors;
	\item wake-up radios (WURs). In scenarios with more energy-demanding devices which cannot rely entirely on RF energy sources, a passive RF EH chip still may be used as a WUR that generates a wake up pulse upon receiving a command from a nearby 
	transmitter \cite{Soyata.2016}. The use of a fully passive WUR means that the active portion of the sensor will only be active for short periods, and does not need to actively listen for commands during downtime, which certainly improves the devices' lifetime.
\end{itemize}
Note that different applications/use cases may use WET services provided by the same RF transmitter, which in turns must address their contracted, potentially heterogeneous, QoS requirements.
%
\begin{figure}[t!]
	\centering
	\includegraphics[width=0.49\textwidth]{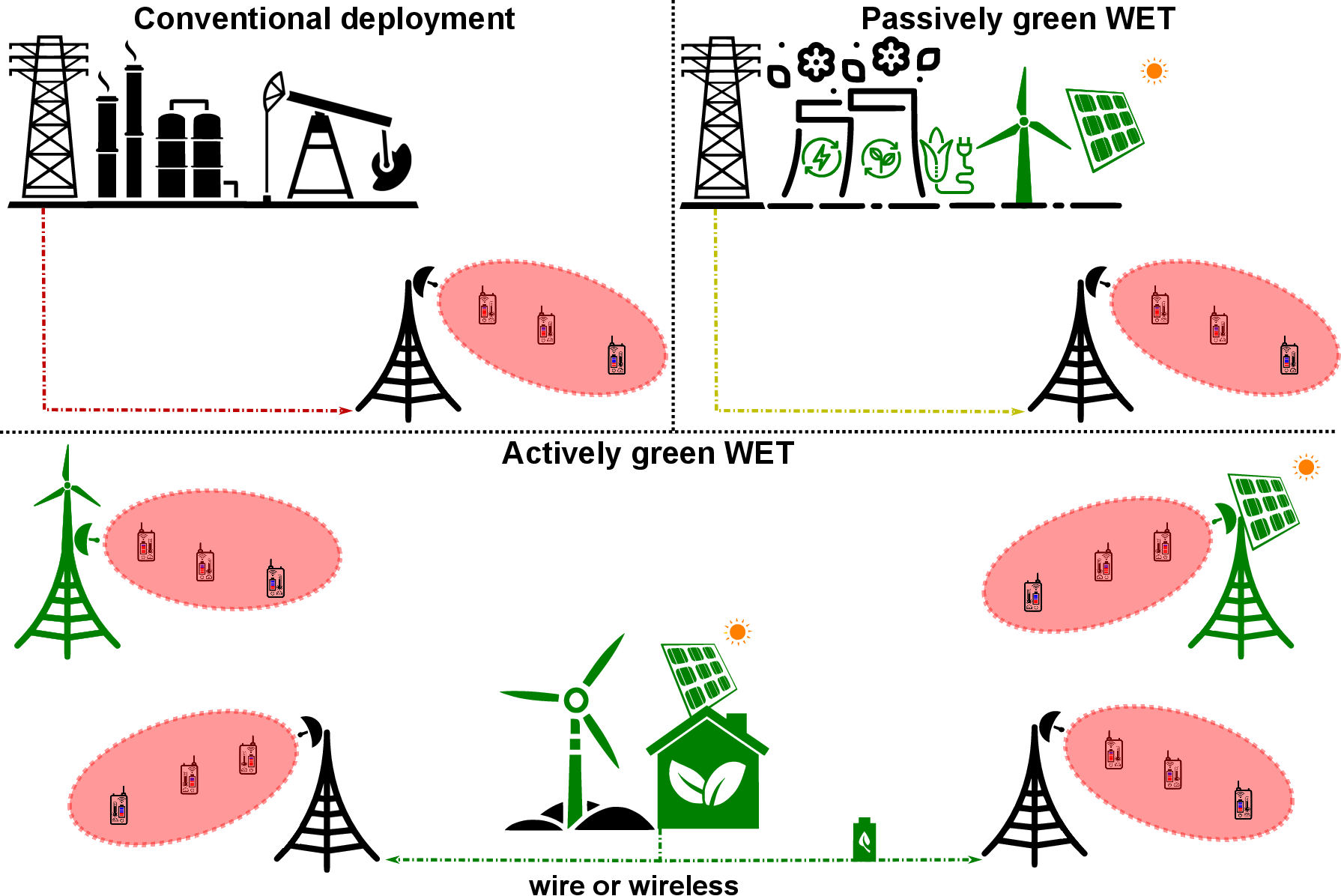}
	\caption{Sustainable WET: from conventional deployments to passively/actively green WET.}	
	\label{greenWet}
\end{figure}
\subsubsection{Green WET}\label{green}
WET is undoubtedly a sustainability promoter.
However, the usual conception of WET-enabled networks, where PBs and hybrid BSs are reliably powered by traditional energy sources, e.g., carbon-based energy, is not strictly compatible with the vision of a fully sustainable IoT.
Instead, WET becomes greener by powering PBs and hybrid PBs using energy harvested from renewable sources \cite{Tran.2018,LiuAnsari.2019}. This can be implemented in two main ways:
\begin{itemize}
	\item passively, where no locally-available green energy sources are exploited. The green energy is harvested at  the industry side and distributed to the WET transmit infrastructure, which requires specific green contracts with the energy provider \cite{Ghesla.2017}; or
	\item actively, where the green energy is harvested locally by the equipment of the WET provider. For example, PBs connected with outdoor solar panels may use such energy to wirelessly recharge indoor RF-EH devices.
\end{itemize}
These variants are illustrated in Fig.~\ref{greenWet}, and note that other dedicated WET technologies, e.g., laser power beaming, can be incorporated and serve as bridge to transport the energy from the green source(s) to PBs.

Among above variants, active green WET is the most attractive as it exploits local renewable sources to strictly realize a small-scale sustainable ecosystem by itself. However, it comes with the typical energy availability challenge of ambient EH. 
To gain in performance robustness against green energy shortages, the aforementioned variants may be hybridly combined, or intelligent energy balancing \cite{LiuAnsari.2019} and trading \cite{Jiang.2019} 
 mechanisms between PBs and hybrid BS may be introduced, e.g., to distribute surplus  energy properly.

Holistic optimization frameworks must consider the fluctuations on the energy availability in the short and large time-scale to fully leverage 
	green WET-related as a sustainable, efficient and scalable technology. In the remaining of the paper, the focus is on the last-segment of such broad sustainable ecosystem: WET from PBs or hybrid BSs to the IoT deployments. Technological enablers for mWET are thoroughly discussed.



\section{Enablers for Efficient and Scalable WET}\label{enablers} 
%
Wirelessly powering the IoT still encounters many challenges ahead, for instance
\begin{itemize}
	\item increasing the end-to-end system efficiency, while limiting further the energy consumption of EH devices \cite{Clerckx.2018};
	\item seamless network-wide integration of wireless communication and energy transfer (at all the system levels) \cite{Clerckx.2018}; 
	\item powering a massive number of devices, while enabling ubiquitous  energy accessibility with QoS guarantees \cite{Lopez.2018_3,Lopez.2020,LopezMahmood.2020}.
\end{itemize}
Several techniques and technological trends that 
seem suitable
for enabling WET as an efficient and competitive solution for sustainably
powering 
future IoT networks are discussed next. 
\subsection{Energy Beamforming (EB)}\label{EB}
\begin{figure}[t!]
	\centering
	\includegraphics[width=0.48\textwidth]{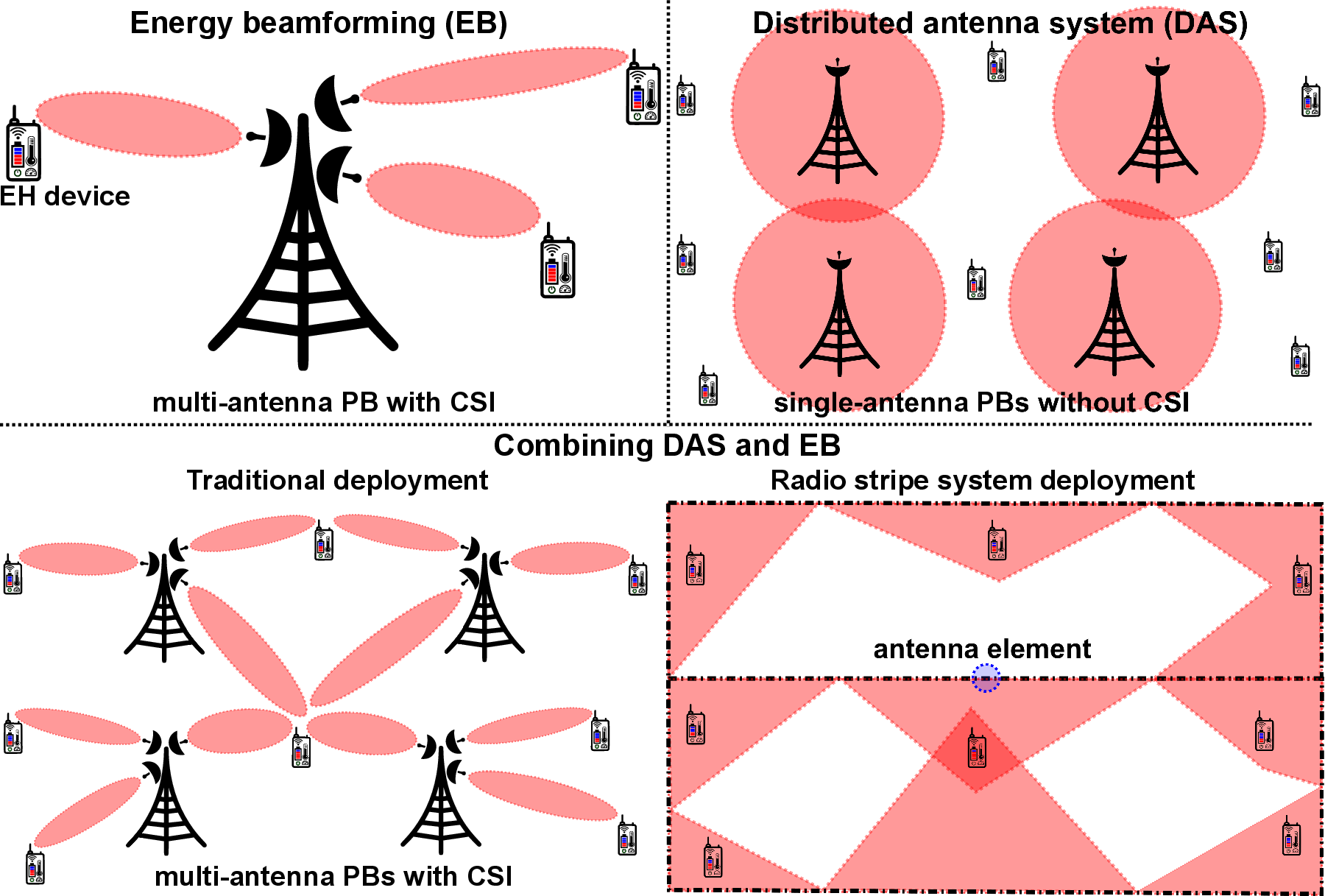}
	\caption{Enabling efficient WET: \textit{a}) EB (left-top), \textit{b}) DAS (right-top) and \textit{c})  DAS \& EB (bottom).}
	\label{FigEBDAS}
\end{figure}

%
 EB allows focusing energy in narrow beams toward the end devices  as shown in Fig.~\ref{FigEBDAS}a, which improves end-to-end efficiency. This is accomplished by carefully weighing the energy signals at different antennas such that a constructive superposition is attained at intended receivers.
The 
larger the number of
 antennas $M$, the sharper the energy beams the PB can generate in some specific spatial directions.
Additionally, $M$ limits the number of beams, thus it is required that $S\le M$ to reach each of the $S$ deployed EH devices with a dedicated energy beam
 \cite{Bi.2016,Clerckx.2018,Lopez.2018_3,Choi.2018}.

Traditional EB requires accurate CSI,
 including both magnitude and phase shift from each of the transmit antennas to each receive antenna. 
 However, in practice, CSI 
is difficult/costly to acquire in WET systems. 
On the one hand, sending training pilots and waiting for a feedback from the EH devices is not desirable since: 
\begin{itemize}
	\item many simple EH devices do not have the required baseband signal processing capability to perform channel estimation \cite{Bi.2015}; 
	\item significant amounts of time and energy are required for achieving accurate channel estimation, which may erase (or even reverse) the gains from EB \cite{Lopez.2020_mag,Zeng.2015,ZengZhang.2015};
	\item the problem of reliable CSI feedback persists \cite{Bi.2015,Lopez.2020_mag}.
\end{itemize}
Alternatively, it seems appropriate that the EH devices send the training pilots, while the PB estimates the CSI and forms the energy beams. This reverse-link training becomes suitable
when using large antenna arrays, e.g., massive MIMO (mMIMO) \cite{Flordelis.2018} or large intelligent surface (LIS) \cite{Wu.2020}, since training overhead is independent of the number of PB's antennas. However, the EH devices still need carefully designed training strategies, such as the transmit power, duration, and frequency bands, to minimize their energy expenditure \cite{Zeng.2015,ZengZhang.2015}.
Notice that 
an efficient MAC orchestration\footnote{In general, a basic time/frequency-division multiple access 
scheme is not appropriate, and  more evolved MAC schemes are required to  account for the different (and variable) energy requirements of the EH devices, and the  QoS	demands of both data transmission and energy transfer in WPCNs \cite{Choi.2019}.} is  necessary to avoid pilot collisions and consequent extra energy expenditure in the collision resolution phase. Additionally, accurate channel reciprocity must hold, even though it is sensitive to hardware impairments specially when devices at both link extremes are very different. For some scenarios, there is also the problem of receiver mobility which could lead to time-varying channels and makes channel tracking difficult. 
In both downlink or uplink training, the energy/time limits CSI acquisition procedures, which potentially produce substantial errors in estimation and quantization. 
Practical designs should take these phenomena into account.
\subsection{Distributed Antenna Systems (DAS)}\label{DAS}
End-to-end efficiency of EH systems decays quickly with distance between the PB and EH device because of the severe power attenuation. 
%
DAS (or distributed PBs)-based solutions as shown in Fig.~\ref{FigEBDAS}b are  undoubtedly
appealing 
to ban blind spots, while homogenizing the energy provided in a given area, and supporting ubiquitous energy access \cite{Bi.2015,Choi.2017,Choi.2018}.
If each separate multi-antenna PB is responsible for powering a smaller set of EH devices as shown in Fig.~\ref{FigEBDAS}c, then CSI acquisition issue is alleviated \cite{Choi.2018}.
Moreover, distributed EB based on local CSI does not require frequency and phase synchronization among the PBs, unlike that based on global CSI for which it is  very costly and challenging.

\begin{figure}[t!]
	\centering
	\includegraphics[width=0.44\textwidth]{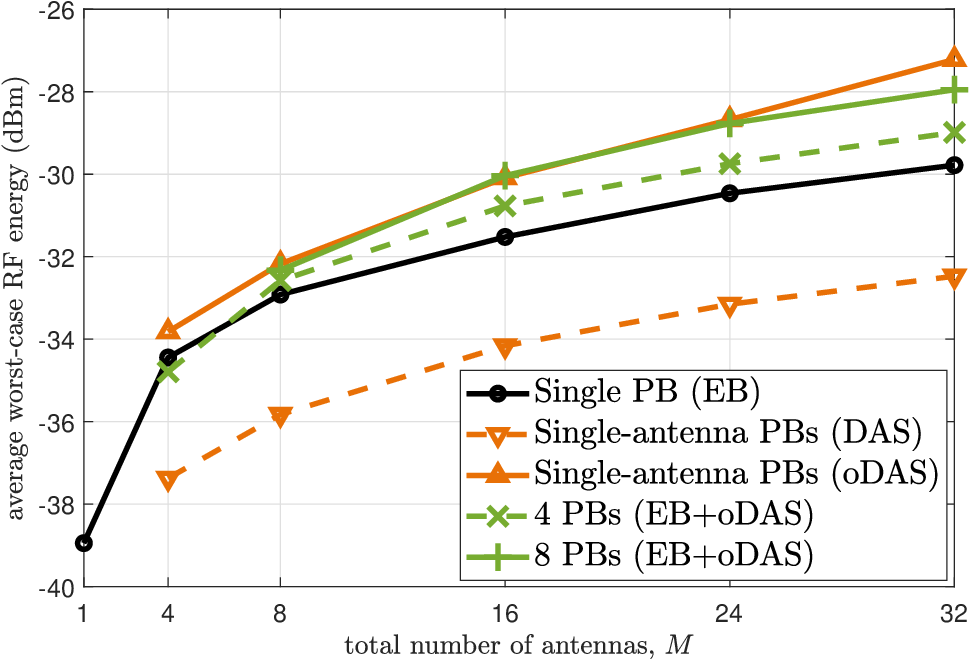}
	\caption{Average worst-case RF energy at the input of the energy harvesters (average max-min performance) as a function of the total number of transmit antennas $M$ for different EB, DAS and hybrid DAS \& EB deployments.
	}	
	\label{figDAS}
\end{figure}

As an illustrative example consider the results in Fig.~\ref{figDAS}, which show the worst-case average RF energy delivered to $S=64$ EH devices deployed randomly in a $20$ m radius area as a function of the total number of transmit antennas $M$. 
Note that the illustrated performance corresponds to the IoT device, among the entire set of harvesters, with the minimum average RF energy available at the EH circuit's input.
The system parameters are given in Table~\ref{table2}. Note that WET channels are usually under a strong line of sight (LOS) influence, thus, the Rician distribution is usually appropriate for fading modeling \cite{Lopez.2018_3,Lopez.2020,LopezMahmood.2020,Lopez.2020_letter,Lopez.2020_mag}. In this case, a LOS factor of 10 dB, i.e., 10 dB above the non-LOS components, is adopted  \cite{Lopez.2020_letter}\footnote{Rician fading channels with LOS factor of 10 dB are approximately equivalent to channels under Nakagami-m fading with factor $m\approx 5.7$.}. Meanwhile, PBs are assumed equipped with uniform linear arrays (ULA) to keep the modeling and analysis simple\footnote{Refer to \cite{Lopez.2020} for the mathematical modeling of Rician fading channels with ULA transmitters.}, and operation is assumed in the ISM 900 MHz band. Additionally, since the system performance scales linearly proportional with the total transmit power, the value of the latter is assumed normalized without loss of generality.  The following PB deployments are considered:
\begin{table}[!t]
	\centering
	\caption{Simulation parameters}
	\begin{tabular}{cc}
		\toprule
		\textbf{Parameter} & \textbf{Value/Model}  \\
		\midrule
		Total transmit power  & $1$ W  \\ 
		Transmit antenna array architecture  & Half-wavelength spaced ULA\\
		Average power attenuation at 1 m  & $30$ dB (typical in ISM 900 MHz)  \\
		Path loss exponent & $2.7$ (typical in short-range setups) \\
		Small-scale fading model & Quasi-static Rician fading\\
		LOS factor $\kappa$ & $10$   \\
		Number of EH devices $S$ & $64$ \\
		EH devices' deployment & Randomly uniform in the area \\
		\bottomrule
	\end{tabular}\label{table2}
\end{table} 
\begin{itemize}
	\item single PB (EB): a PB located at the circle center equipped with $M$ antennas;
	\item single-antenna PBs (DAS): $M$ single-antenna PBs random and uniformly distributed in the area;
	\item single-antenna PBs (optimized DAS --oDAS): similar to the single-antenna PBs deployment but the PBs' locations are set using the well-known \textit{K-means} clustering algorithm \cite{Bishop.2006};
	\item 4 PBs (EB+oDAS): four PBs, each equipped with $M/4$ antennas, are deployed and their locations are set using the \textit{K-means} clustering algorithm;
	\item 8 PBs (EB+oDAS): eight PBs, each equipped with $M/8$ antennas, are deployed and their locations are set using the \textit{K-means} clustering algorithm.
\end{itemize}
Additionally, the transmit power of each individual PB in the 4/8 PBs configuration is set proportional to the path loss of its farthest EH device, while PBs' sum transmit power is constrained to 1 W. Each multi-antenna PB uses EB to reach its associated EH devices with maximum fairness, i.e., no device is expected to benefit more from PB's WET than others. 
This problem, although not convex in general,  can be transformed to a semi-definite program (SDP) as in \cite{Thudugalage.2016}, which in turn can be solved efficiently. Note that a naive DAS, without properly setting the PBs locations, performs poorly, and even a single centered PB is preferable. 
By optimizing the PBs locations, significant performance gains can be attained. However, this may not be feasible in more dynamic setups where the EH devices are not completely static and/or demand time-varying heterogeneous QoS. Note also that the deployment costs increase as the number of PBs increases. In such scenario, and for the same total number of transmit antennas $M$, it is preferable deploying multi-antenna PBs and take advantage of both DAS and EB. The latter is realized even with  less CSI acquisition overhead and associated energy expenditures compared to the single multi-antenna PB implementation.
The greater the total number of antennas, the more multi-antenna PBs should be deployed for the best system performance. 
As evidenced, optimizing PBs  (or hybrid BSs) placements is key to ensure DAS benefits in WET-enabled networks, and it has been considered in different scenarios, e.g., \cite{BiZhang.2016,Dai.2018,Osmel.2020}.
Still, much more research effort is required towards efficiently designed DAS for massive WET. In such scenario, traditional clustering algorithms may be strictly sub-optimal since a PB powering certain cluster may significantly influence other nearby clusters as well. Novel PB positioning techniques, with/without EH devices' position information\footnote{Note that deploying PBs without exploiting EH devices' position information may be needed in some scenarios to ensure area-wide energy availability, thus, supporting devices' mobility and long-term network/environment dynamics \cite{Osmel.2020}.}, considering the network-wide effect of each PB, are required, and constitute an open research direction.
Moreover, additional performance improvements can be attained via proper multi-antenna PB rotation  \cite{Lopez.2020_letter}, which influences the LOS channel. PB rotation can be accomplished either by the technician in static setups or by equipping the PB with a rotary-motor in more dynamic setups as discussed in the next subsection. %

Finally, future WET systems will benefit from novel kinds of DAS deployments such as cost-efficient radio stripe systems \cite{Interdonato.2019}. In these systems, actual PBs will consist of antenna elements and circuit-mounted chips inside the protective casing of a cable/stripe. While traditional antenna deployments may be bulky, radio stripes enable 
imperceptible
installation
and alleviate the problem of deployment permissions. Besides, cables are malleable, and the overall system is resilient to failures because of its distributed functionality. Optimized resource allocation schemes, circuit implementations, prototypes, and efficient distributed processing architectures to avoid costly signaling between the antenna elements, are needed.
\subsection{Enhancements in Hardware and Medium} \label{HW}
\subsubsection{Ultra-low-power Receivers}
Powering a certain IoT device via WET becomes easier as its energy demands become less stringent. Ultra-low-power receive architectures are thus essential to fully realize the potential of WET in large-scale IoT deployments. EH hardware optimization has traditionally focused either on the antenna design (for reduced form factor and/or high antenna gain) \cite{Divakaran.2019,Mathur.2018,Nguyen.2018}, matching network \cite{Divakaran.2019,Merz.2016,Hameed.2017}, rectifier and rectennas (from single to multi/tunable-band, and from frequency-selective to wide-band) \cite{Divakaran.2019,Tran.2017,Yaldi.2016,Chaour.2017,Colaiuda.2020}, voltage multiplier \cite{Divakaran.2019,Rajawat.2018,Azawy.2019,Majdi.2017}, or even the power management unit (PMU) \cite{Popovic.2014,Ababneh.2017,Cansiz.2019}.
The PMU is nowadays an optional circuit block in EH circuits, which tracks and optimizes the EH efficiency. Specifically, the PMU monitors the harvested energy levels and provides the charge control/protection of the energy storage units such as capacitors or batteries \cite{Cansiz.2019}. However, the intelligence that can be deployed into a PMU is limited by its allowed energy consumption. In general, significant power reduction/efficiency gains can still be achieved in the coming years by optimizing the EH hardware as a whole, e.g., by integrating antennas/rectennas into the device package with an optimized intimate connection \cite{Mahmood.2020}, and designing/integrating PMUs that optimize the net harvested energy (harvested energy minus power consumption over time) to ensure real system gains.
Additionally, WURs, duty cycling or event-driven architectures (with short power-up settling times to enable swift change between sleep and on states) may be adopted to reduce transceiver usage. 

Future ultra-low-power architectures face the additional challenge of seamless circuit integration into everyday object/materials, on which there are some recent advances. 
For instance,  a prototype of an  optically transparent (microstrip patch) EH antenna on glass was proposed in \cite{Vullers.2008}. The circuit can be quasi-invisibly installed in a window without compromising its function. 
Meanwhile, authors of \cite{Vital.2020} demonstrate flexible and lightweight textile-integrated rectenna arrays for powering wearable electronic devices by exploiting large clothing-areas. A textile antenna for wearable RF-EH in the sub-mmWave band was also proposed in \cite{Wagih.2019}.
Additionally, technologies such as  printing
and roll-to-roll compatible techniques that enable  the integration of EH electronics on large surfaces are discussed in \cite{Roselli.2014}. 
Future advances on ultra-low-power circuits and their integration into a plethora of materials will significantly broaden the horizons of WET technology in the 6G era.
\subsubsection{Enhanced PB}\label{enhanced}
\begin{figure}[t!]
	\centering
	\includegraphics[width=0.49\textwidth]{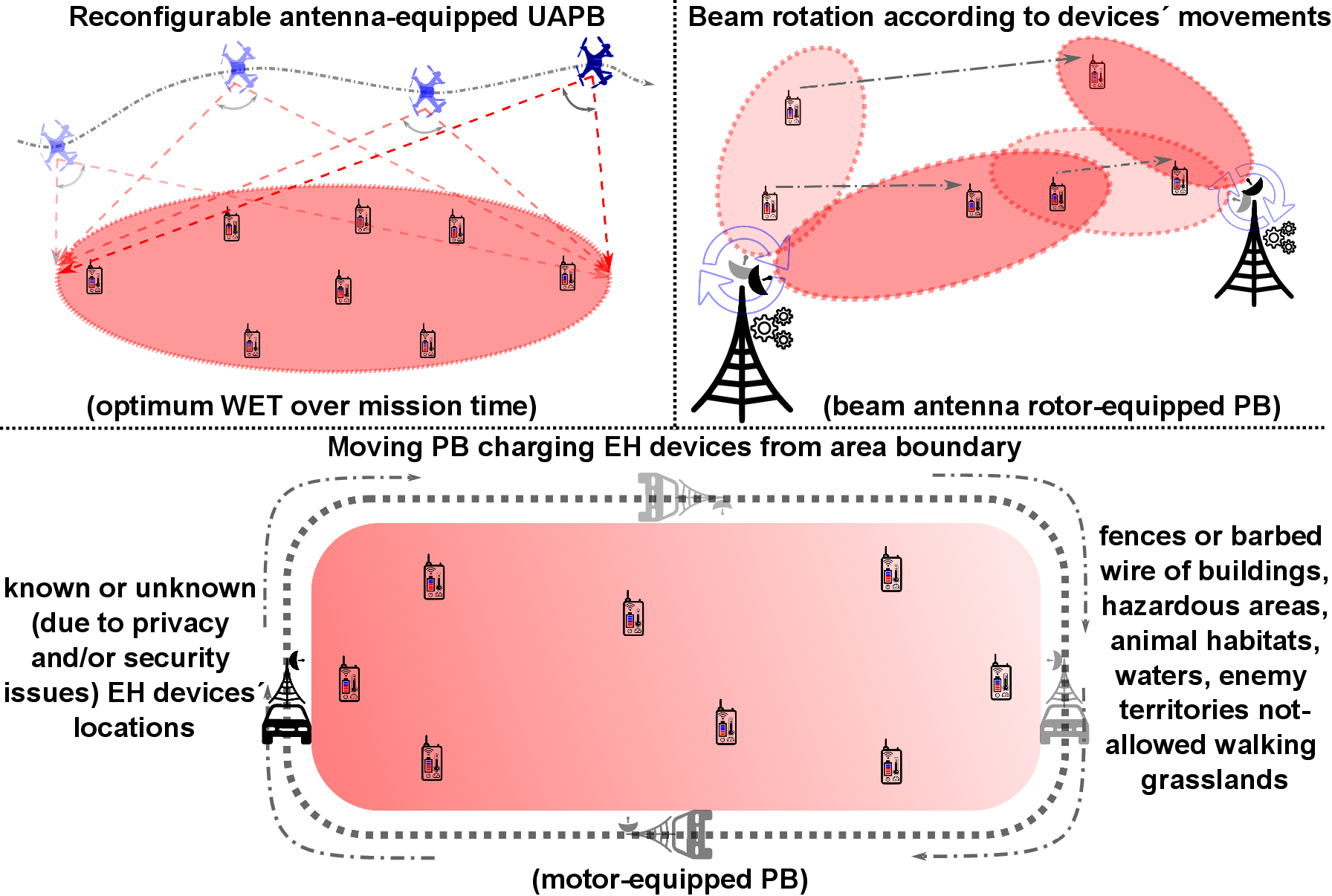}
	\caption{Enabling efficient WET with enhanced PBs: UAPBs and rotor/motor-equipped PBs.}	
	\label{UAPB}
\end{figure}

The traditional fixed PB concept is evolving to more efficient and flexible implementations as illustrated in Fig.~\ref{UAPB}. The radio stripe-based PB discussed in Section~\ref{DAS} constitutes an example of an efficient  architecture that allows imperceptible installation and potentially wide coverage. Also, the increasing adoption and miniaturization of massive antenna arrays and high-gain antennas is promising in terms of transmit hardware. As for hardware adaptability, 
early results in \cite{Lopez.2020,Lopez.2020_letter} suggest for instance that
a rotor-equipped PB with optimized rotation may enable local mWET. However, how to optimally and dynamically rotate the PB in both single and multi-PB scenario remains as an open problem.
Additionally, motor-equipped PBs for mobile charging is gaining attention in recent years, e.g., \cite{Dai.2014,Zhang.2018,Wang.2019,Dai.2020}.
Such mobile PBs are designed to periodically or on-command wander around their service area to recharge full/partially the corresponding set of EH IoT devices. The advantages of mobile charging with respect to static PB deployments include, but are not limited to \cite{Dai.2020}: i) fewer required PBs  and ii) more adaptability to a dynamic IoT network topology 
as charging routes can be easily re-adjusted by a centralized network planner entity or even autonomously by the PB itself via AI/ML
mechanisms. 
By incorporating high-gain directional antennas into the PBs, further considerable improvements can be attained, as illustrated in \cite{Chang.2015,Moraes.2017}. 
However, accurate pointing toward the receivers is necessary and may require the positioning service offered by beyond 5G/6G networks in dynamic setups \cite{Mahmood.2020,Ghosh.2019}.

Meanwhile, PBs mounted on UAVs may provide additional advantages over ground PBs such as i) aerial mobility, ii) flexible accessibility to remote, rural or disaster areas where gaining ground access is difficult/impossible \cite{LiuDai.2020}, and iii) great chances of reaching positions with LOS channels to the IoT devices for downlink WET and downlink/uplink WIT. Therefore, unmanned aerial PBs (UAPBs) can potentially foster RF-EH IoT applications such as  environment/farm monitoring and emergency services, where UAPBs wake up IoT deployments via WET and collect data, as illustrated in Fig.~\ref{Fig1}. Although there is vast literature on performance optimization of UAPBs-enabled IoT scenarios, e.g., \cite{Xu.2018,Suman.2018,LiuDaiWang.2020}, still further efforts are needed to overcome the UAPBs' limited powering range, energy efficiency issues and optimize the flying trajectory design \cite{LiuDai.2020}. Further advances on reconfigurable antennas \cite{Hussaini.2015,Wolfe.2018} and UAV swarms operation constitute potential technological enablers \cite{Zeng.2018} as they allow realizing proper beam footprint adjustments and high transmit gains, which may significantly boost the QoS-based energy coverage of the UAPBs.

\subsubsection{Reprogrammable Medium}
\begin{figure}[t!]
	\centering
	\includegraphics[width=0.48\textwidth]{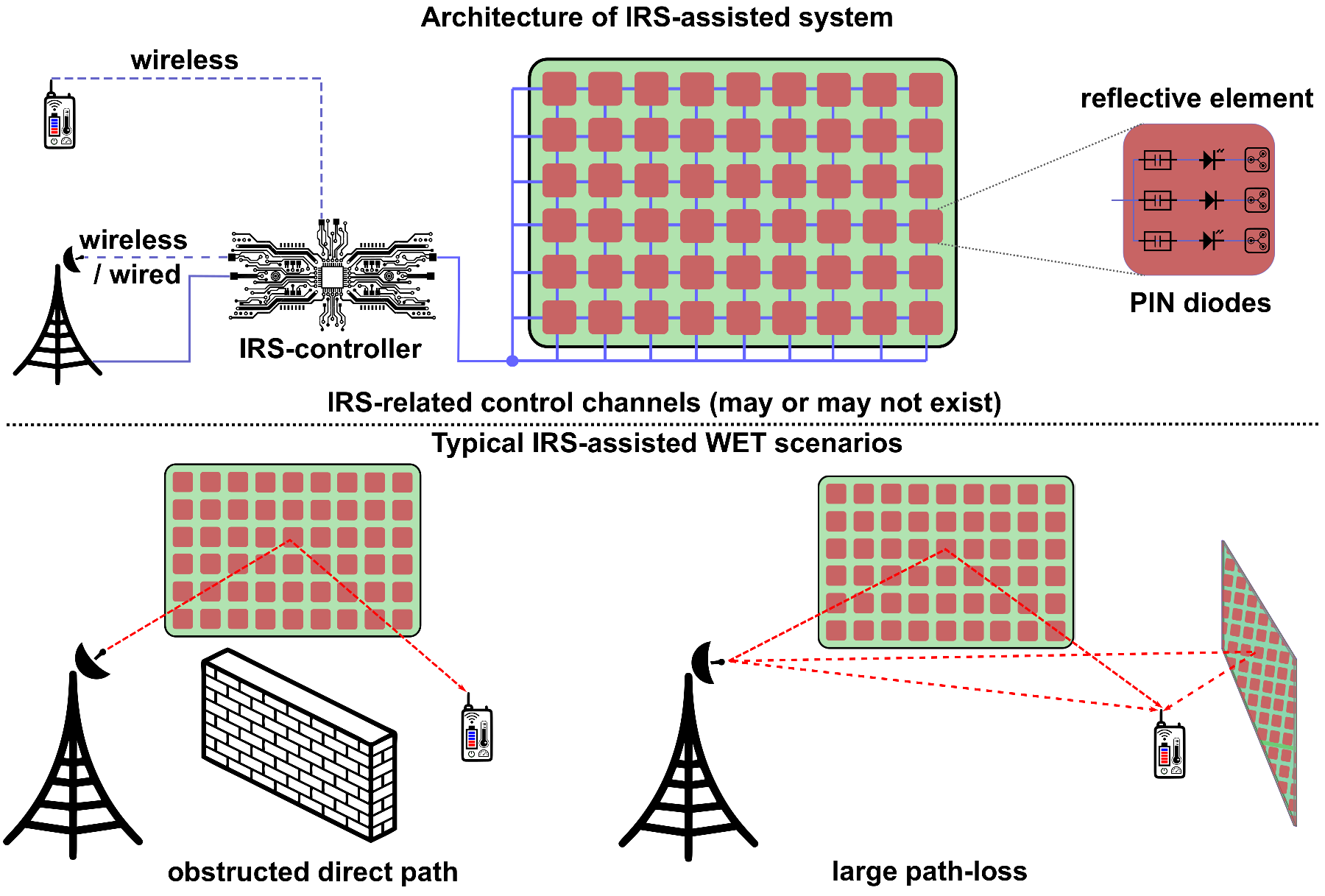}
	\caption{IRS enabling efficient WET: $a)$ Architecture of IRS system (top) and $b)$ typical IRS-assisted WET scenarios (bottom). Note than in case no control channels are established with the IRS, then, PB and user must establish one.}
	\label{IRS}
\end{figure}

In addition to transmitter/ receiver-related enhancements, the WET propagation medium can also be conveniently `influenced' via the strategic deployment of smart reflect-arrays and reconfigurable meta-surfaces \cite{Renzo.2019}.  
For instance, intelligent reflecting surfaces (IRS) allow on-fly and opportunistic reconfiguration of the propagation environment without additional energy consumption/expenditure, and are considered a key technological enabler of future green networks \cite{Renzo.2019,WuZhang.2020}. 
The architecture of an IRS-assisted system is illustrated in Fig.~\ref{IRS}a. Observe that IRS are composed of a large number of low-cost and passive reflective components (controlled by positive-intrinsic-negative (PIN)
diodes, field-effect transistors, or micro-electromechanical switches) able to dynamically shift the phase of incident RF signals  to ensure they are added either constructively, or even destructively to limit interference, at a target receiver.

Research community has mainly focused on IRS-assisted communication scenarios, but IRS-assisted RF-powered systems has been gaining interest recently, e.g., refer to  \cite{Lyu.2020} for an analysis and optimization of an IRS-assisted WPCN, while an IRS-assisted SWIPT setup is investigated in \cite{Tang.2020,Pan.2020,WuZhang2.2020}.
Particularly attractive are the analysis and discussions carried out in \cite{Pan.2020,WuZhang2.2020}, which are general enough to be extrapolated to purely WET-setups (the focus of our work) since EH devices and information decoding devices are considered as separate entities. Therein, authors show that the IRS deployment allows important power saving at the hybrid BS/PB serving a QoS-constrained network.

In general, IRS are beneficial to either compensate the significant path losses thanks to their large aperture arrays, or to provide alternative LOS WET links to obstructed direct paths as illustrated in Fig.~\ref{IRS}b. Towards the massive adoption of this technology, there are still some design trade-offs and challenges that need to be carefully addressed, for instance \cite{WuZhang.2020}:
\begin{itemize}
	\item manufacturing high-precision reflective elements requires  expensive 	hardware, which may not be a scalable solution
	as the number of elements becomes very large. For example, to enable $2^v$ levels of phase shifts, a number of $v$ PIN diodes (with output states `on' and `off') needs to be integrated to each reflective element. Under form-factor constraints, such hardware design becomes extremely challenging, also because more controlling pins from the IRS controlled are needed;
	\item the passive reflect EB of IRS and active EB of PBs and hybrid BS need to be jointly designed for optimum performance, which leads to complicated optimization problems that are hard to be efficiently solved. Also, different from digital or hybrid digital/analog active  EB, which natively allows dealing with frequency-selective channel variation, the passive EB design at the IRS  becomes further challenging when aiming at jointly balancing the channels at different frequency sub-bands;
	\item similar to traditional EB (discussed in Section~\ref{EB}), instantaneous CSI is required to leverage the passive EB performance gains promised so far in the literature. 
	CSI may be acquired either by i) equipping each or a cluster of reflective elements with a low-power receive RF chain to enable channel estimation, or ii) estimating the composite transmit and IRS-assisted channel with IRS reflection patterns known a priori, or exploiting receivers' feedback without estimating channels. However, in IRS-assisted WET to a large number of low-power IoT devices, such overhead may be prohibitive, and efficient CSI-limited/free schemes need to be leveraged.
\end{itemize}
\subsection{New Spectrum Opportunities}\label{spectrum}
Recently, the electromagnetic spectrum from $28-300$ GHz is being considered by research community, and even starting to be exploited by industry, for wireless communication applications. This is motivated by \cite{Niyato.2017,Nawaz.2020}:
	\begin{itemize}
		\item the large bandwidths that remain unexploited in such spectrum region and the scarcity of the spectrum currently under widespread usage;
		\item the propagation of signals in the mmWave frequencies is more directive and of shorter-range, which is favorable for spectrum re-use in small cells; and
		\item shorter wavelengths allow reducing antenna sizes, which translates to either smaller form-factors e.g., at the IoT device side, or to being able of packing more antennas, e.g., at the BS side, and attain high directional gains.
	\end{itemize}
From the WET point of view, above advantages hold and align with the vision of massive miniaturized IoT devices' deployments, e.g., smart dust like motes and zero-energy sensors \cite{Portilla.2019}, in an ultra densely connected world.
Also, an easier network integration/coexistence of WET and WIT may be promoted since interference issues are considerably relaxed.
Furthermore, the larger spectrum bandwidth\footnote{Note that modulated mmWave WET demands the receiver to be equipped with an ultra-wideband RF-EH circuit, e.g., as in \cite{Wagih.2019}.} and high antenna gains may allow the PB to effectively transfer larger amounts of energy under LOS conditions. Since LOS channels are characteristic of many WET applications and are mostly influenced by the antenna array topology and network geometry \cite{Hampton.2014},  CSI-based EB may be avoided (at least partially) by exploiting accurate devices' positioning information. However, the problem of imperfect beam alignment must be carefully taken into account since it is known to significantly degrade mmWave WET performance \cite{Wang.2020}.
How to overcome non-LOS, or even
severe signal attenuation in rainy conditions \cite{Kamga.2019}, is an even more critical challenge  in mmWave WET with limited CSI, and may require exploiting multi/hybrid radio access technologies. 
Finally, the highly directive  energy beams that are typical in mmWave WET may cause the signal intensity perceived in a particular area to be strong enough to harm human health\footnote{The radiation power of any device at certain frequency must satisfy an equivalent isotropically radiated power requirement \cite{Bi.2015}.}. DAS (discussed in Section~\ref{DAS}) is a promising approach to solve safety issues as smaller path losses need to be compensated by the beam gains, and could be even combined with sensing-based humans real-time detection technologies to cease beam-specific WET when it deems to be harmful.
\subsection{Resource Scheduling and Optimization}\label{resource}
Under intelligent policies, WPCN may achieve a performance, e.g., in terms of throughput \cite{Ju.2014}, comparable to that of a conventional non-WET network. When WIT and WET take place in the same network, scheduling should i) avoid co-channel interference or limit its impact (e.g., by taking advantage of synchronization and SIC techniques to cancel deterministic WET signals in the network as exemplified in Fig.~\ref{Fig2}), and ii) optimize the overall system performance according to the metric of interest.
Novel fine-grained performance metrics based on meta distribution (e.g., of the SINR, transmit rate, or even harvested energy) \cite{Deng.2019}, which are suitable for resource allocation supporting per-link service guarantees, may be extremely useful  to address the extreme QoS demands of future 6G systems.
In practice, real-time information/energy scheduling is a challenging problem because of time-varying wireless channels and the causal relationship between current WET process and future WIT.  

Multi-antenna energy and information transmitters endow spatial domain communication and energy scheduling.
For instance, PB utilizes EB to steer stronger energy beams to efficiently reach certain users while prioritizing their energy demands toward the information transmission phase. Besides, combination of EB, space-division multiple access, and dynamic time-frequency resource allocation further enhances system performance in WPCNs. 
NOMA techniques have also been recently  considered as a performance booster in WPCNs, e.g., \cite{Liu.2019,Liu.2020}. However, more in-depth studies are still needed to  optimize the performance and assess the suitability of NOMA under the degenerative effect of a higher circuit power consumption and additional overhead, e.g., for CSI-acquisition and user scheduling, that may demand its adoption in practice \cite{Wu.2018,Chi.2019}.

Some other strategies considered in the literature are input signal distribution optimization 
\cite{Clerckx.2018,Hu.2019},
cooperation \cite{Alsaba.2018,Lopez.2017_2,Lopez.2018_4}, hybrid automatic repeat-request 
\cite{Witt.2014,Makki.2016}, power control \cite{Clerckx.2018,Lopez.2018_2,Lopez.2020_2,WuZhang2.2020} and rate allocation \cite{Lopez.2018_1,Lyu.2020,Pan.2020}.
However, the heterogeneous traffic and QoS requirements of future IoT deployments demand adaptive use-case tailored solutions \cite{Ramezani.2017}, which may be assessed by adopting AI/ML mechanisms allowing autonomous re-configuration to network/channel/requirements variations. Such network intelligence may inherently deal with the many non-linear impairments that affect the EH hardware, which must be considered but are difficult to model analytically. In the context of WET-enabled systems, AI/ML optimization frameworks have been mainly proposed to be deployed at the PB or hybrid BS side, e.g., \cite{Kang.2018,Iqbal.2019,Hameed.2020}, since embedding the required algorithmic functionalities on chip remains a challenge under the strict limitations imposed by the IoT device's power consumption. Toward future 6G systems, on-chip intelligence needs to be developed to run/facilitate at least small-scale tasks, e.g. smart-wake up \cite{Mahmood.2020} and PMU functionalities.
Finally, note that computation tasks that remain too costly, either related to scheduling and optimization, or sensed data processing,  may be offloaded to a mobile edge computing (MEC) server for further processing and energy saving \cite{HuWong.2018,LiuXu.2019,Li.2020}. 	
\subsection{Distributed Ledger Technology (DLT)} 
WET, as a key component of RF-EH networks, is fundamental for realizing the Internet of Energy 
paradigm, which also includes energy trading in microgrids and vehicle-to-grid networks
\cite{Li.2018}. 
Both infrastructure and IoT nodes may trade their energy goods or surplus with other nearby nodes for which they can act as opportunistic PBs. The distributed nature of this trading creates important challenges in terms of security and privacy, e.g., 
potential attacks from malicious IoT devices such as energy repudiation of reception and energy state forgery \cite{Jiang.2019}.
DLT-based solutions, e.g., blockchain and holochain DLT, are promising technologies
 since they allow value transactions between parties through decentralized trust, although key
 challenges such as large communication overhead, handling massive two-way connections and energy-efficient DLT protocol design, 
 still 
  need to be efficiently addressed. Moreover, powering the devices that actually \textit{paid} the service must be extremely precise in space (e.g., by exploiting narrow EBs), time, and other domains, for which novel and efficient EB designs are necessary, e.g., to guarantee the agreed QoS levels of the legitimate EH devices with minimum communication overhead.
%
%
%
\section{CSI-limited EB Schemes for mWET}\label{scenarios}
%
%
As highlighted in the previous section, the problem of CSI acquisition in multi-user WET systems is critical and limits the practical significance of any research work relying on perfect CSI.
The problem escalates with the number of EH devices. To prevent interference and collisions during training, scheduling strategies may be necessary, draining important energy resources that are costly for energy-limited devices. 
As the network grows, more training pilots may be needed to acquire the whole network CSI within every coherence time interval. The more training pilots used for CSI acquisition, the more stringent the energy demands, thus, traditional training becomes 	inefficient/unaffordable in low-power massive IoT deployments. This problem can be alleviated a bit by noticing that wireless powering efficiency experiences a power-law decay with the distance, thus, the channels from those EH devices farthest from the PB(s) are expected to dominate the beamforming design \cite{Lopez.2020_mag}. Then, channel training can be reduced by avoiding running CSI acquisition procedures for those users that are closer to the PB(s) as illustrated in \cite{Lopez.2020_mag}. However, the farthest, more energy-limited, users would 	still require to spend valuable energy resources for channel training. 
Furthermore, the performance of CSI-based systems decays quickly as the number of users increases \cite{Lopez.2018_3}. Therefore, intelligently exploiting the broadcast nature of wireless transmissions in such massive deployment scenarios is of paramount importance, even more so when powering a massive number of devices simultaneously with minimum or non CSI. 

The following subsections focus on
some approaches that alleviate the need of instantaneous CSI at a multi-antenna PB that powers a massive number of EH devices, including a novel CSI-free scheme that exploits only the devices' clustering information. 
The system parameters given in Table~\ref{table2} are adopted in the numerical examples unless stated otherwise.
%
\subsection{Partial CSI-based EB}
Instead of EB based on instantaneous CSI, the PB may exploit 
 statistical knowledge of the channel, 
 so-called partial/statistical CSI.
 This is three-fold advantageous \cite{Lopez.2020_letter}: 
 \begin{itemize}
 	\item such information varies over a much larger time scale and does not require frequent CSI updates;
 	\item it is learned with limited CSI-acquisition overhead and energy expenditure;
 	\item it is less prone to estimation errors.
 \end{itemize}
%
Partial CSI-based EB schemes are particularly beneficial in setups where the PB is a typical mMIMO or LIS node and downlink/uplink channels are not reciprocal \cite{Qiu.2018,Lopez.2020_mag}. The benefits are surely promising in future network deployments with IRS, since IRS may assist blocked energy transmissions by providing alternative passively controlled LOS paths \cite{Lopez.2020_mag}. CSI estimation/acquisition in this kind of deployments remains a challenge,
while
 partial CSI-based EB could attain near optimum performance as certain LOS is expected between the transmitter and IRS, and between the IRS and the EH devices, which strongly reduce channel uncertainties.
 Still, powering a large number of EH devices remains a challenge since these IRS are passive. One possible approach  may rely on partitioning the reflecting elements set such that each partition is responsible for efficiently reflecting the arriving signal to a certain EH device.

\begin{figure}[t!]
	\centering
	\includegraphics[width=0.44\textwidth]{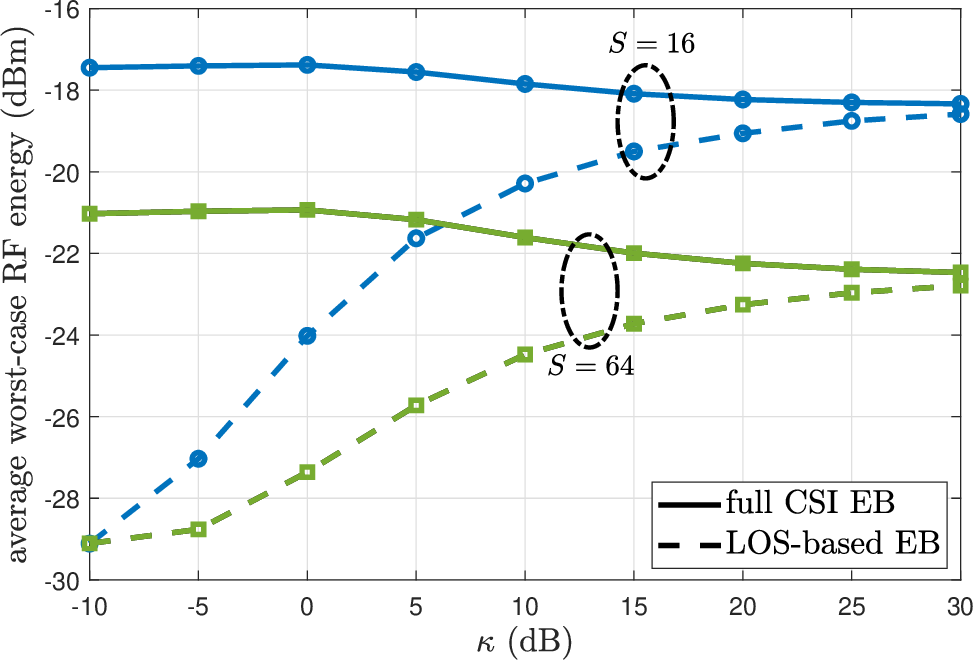}
	\caption{Average worst-case RF energy at the input of the energy harvesters (average max-min performance) as a function of the Rician LOS factor. A PB equipped with $M=32$ antennas powers a set of $S\in\{16,64\}$ EH devices.
	}		
	\label{Fig_r1}
\end{figure}

An EB based on statistical CSI optimizes average performance statistics, e.g., the average harvested energy. For instance, 
assume a simple scenario in which a PB equipped with $M=32$ antennas  powers a large set of 
 EH devices deployed in a $10$ m radius area. Similar to the example discussed in Section~\ref{DAS}, a max-min EB is adopted here. Supposing the PB only knows the first order channel statistics, e.g., the LOS channel in Rician fading, the same solving procedure using the SDP formulation may be applied for the LOS-based
   EB \cite{Lopez.2020_letter}. Fig.~\ref{Fig_r1} shows the average worst-case RF energy available at the EH devices. 
The performance of the statistical EB improves as the LOS
becomes stronger. 
This is 
extremely 
relevant for WET systems that, due to short separation distances, often operate under a strong LOS, e.g., $\kappa\gtrsim 10$ dB.
Exploiting second-order statistics of the channel, such as the covariance matrix, can enhance the performance even more, though acquiring such information may demand more frequent channel sampling. 
All in all, the key finding is that an appropriate statistical precoder may reach near-optimum performance in WET systems.
%
\begin{table*}[!t]
	\centering
	\caption{Comparison of the different state-of-the-art multi-antenna CSI-free WET schemes}
	\begin{tabular}{			
			L{0.34\textwidth} L{0.31\textwidth} 
			L{0.27\textwidth}}
		\toprule \\[-4mm]
		\textbf{Scheme} --\textbf{Description} & \textbf{Massive Support}
		& 
		\textbf{Fitness to DAS/DLT/IRS/mmWave}\\[-0.8mm]
		\midrule
		Antenna phase shifting (APS) with energy modulation/waveform (EMW) \cite{ClerckxKim.2018} --Antennas transmit phase-shifted signals  to induce fast-fading channels.  EMW is introduced to provide
		further enhancements. The non-linearity of the EH circuitry  is exploited. &
		Yes, but for ultra-low power EH scenarios. Limited support to massive deployments of typical EH devices. 
		\emph{Average gain (AG): 	$\sim 1$, 
			Diversity gain (DG):} $\le 2$ (only APS)	
		&  
		Easy integration to DAS but the interference  to coexisting/neighboring networks can be a limiting factor. 
		Not suitable for DLT/IRS/mmWave-based solutions since the energy is omnidirectionally transferred.
		\\ 
		\hdashline \\[-3mm]
		All antennas transmitting independent signals ($\mathrm{AA-IS}$) \cite{Lopez.2020}. & Yes, but preferable for powering devices close and uniformly distributed around the PB. 
		\emph{AG: 	$\!\sim\! 1$, 
			DG: $\!M$}  
		& \vspace{-8mm}
		\multirow{2}{0.27\textwidth}{Easy integration to DAS but the interference  to coexisting/neighboring networks can be a limiting factor (but less than with APS-EMW). 
			Not suitable for DLT/IRS/mmWave-based solutions since the energy is omnidirectionally transferred.} \\[0mm]
		\cdashline{1-2}
		\\[-2mm]Switching antennas ($\mathrm{SA}$) \cite{Lopez.2018_3} --A signal is transmitted  with full power
		by one antenna at a time such that all antennas are used during
		a channel coherence block. & Yes, but preferable for powering devices relatively far and uniformly distributed around the PB. 
		\emph{AG: 	$\!\sim\! 1$, 
			DG:} $M$
		&   \\
		\hdashline
		All antennas transmitting the same signal ($\mathrm{AA}$) \cite{Lopez.2018_3}  or AA with minimum dispersion energy ($\mathrm{AA\!-\!SS}_\mathrm{min\ \!var}$) \cite{Lopez.2020} --The same signal is simultaneously transmitted through all antennas with equal power. & Yes, but  for devices  approximately deployed at the boresight of the PB's antenna and under some LOS. Thus, a maximum of two specifically deployed clusters are supported when using ULAs. 	
		\emph{AG: 	$\le M$, 
			DG:} $\le M$
		& \vspace{-15mm}
		\multirow{2}{0.27\textwidth}{Integration to DAS requires the PBs to carefully coordinate their serving areas, i.e., nearest PB association rule may be highly sub-optimal.
			PBs proper rotations may be needed, and still certain devices may be difficult to reach without the help of strategically deployed IRS. Not suitable for DAS in dynamic setups.
			Their application to DLT-based solutions is possible but limited, while mmWave WET requires  accurate devices' position information and error models.
		} \\
		\cdashline{1-2}
		AA with maximum  average energy ($\mathrm{AA\!-\!SS}_\mathrm{max\ \!E}$) \cite{Lopez.2020} --Similar to $\mathrm{AA}$ and $\mathrm{AA\!-\!SS}_\mathrm{min\ var}$ but the signals in consecutive antennas are shifted $\pi$ radians relative to each other.& Yes, but  for devices  approximately deployed at $90^\circ$ (to the right and left in ULAs) of the PB's antenna boresight and under some LOS. Thus, a maximum of two specifically deployed clusters are supported when using ULAs. The beams are wider than in $\mathrm{AA}$, $\mathrm{AA\!-\!SS}_\mathrm{min\ var}$. 
		\emph{AG: 	$\le M$, DG:} $\le M$ 
		&  \\
		\bottomrule		
	\end{tabular}\label{table}
	\begin{flushleft}
		\footnotesize $M$ is the number of PB's antennas.
	\end{flushleft}
\end{table*}
\subsection{CSI-free Schemes}\label{csifree}
\vspace{-1mm}
Although statistical EB may be viable, some energy is still needed to learn the main channel features, thus it may still be unaffordable in extremely low-power applications,
urging CSI-free schemes \cite{ClerckxKim.2018,Lopez.2018_3,Lopez.2020}.
Such state-of-the-art methods are summarized in Table~\ref{table}. Therein it is also highlighted their fit to support massive IoT deployments served by a single PB, and to adapt to DAS and DLT/IRS/mmWave-based solutions for extending such massive support to multi-PB setups with/without on-demand opportunistic PBs, mmWave WET and IRS-assistance.

\begin{figure}[t!]
	\centering
	\includegraphics[width=0.102\textwidth]{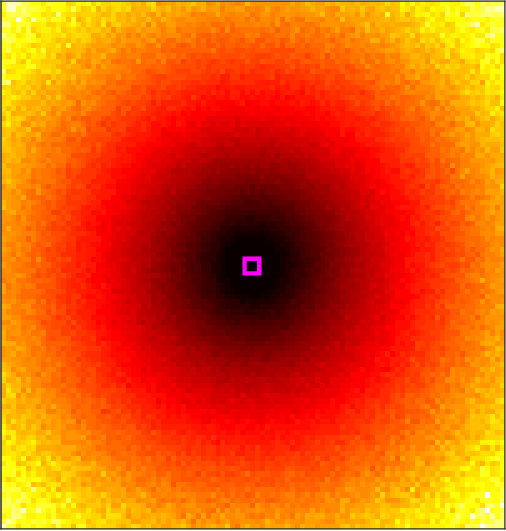}\ \ \ \! \!
	\includegraphics[width=0.102\textwidth]{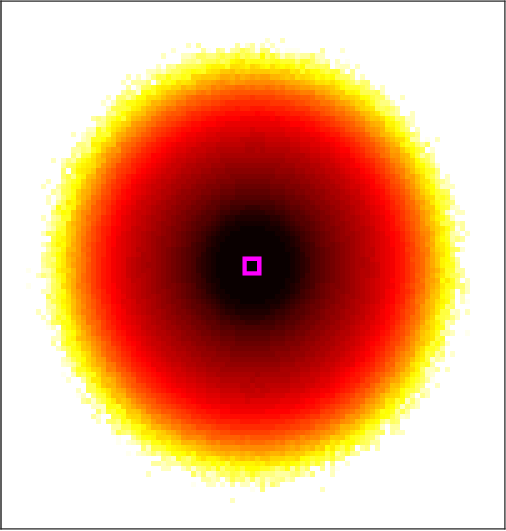}\ \  \ \!\!
	\includegraphics[width=0.102\textwidth]{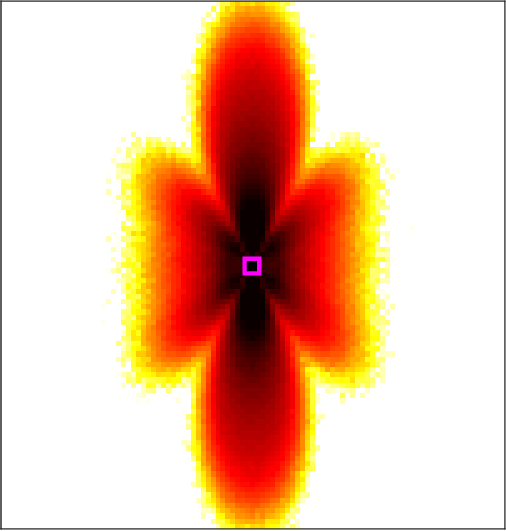}\ \ \ \!\!
	\includegraphics[width=0.121\textwidth]{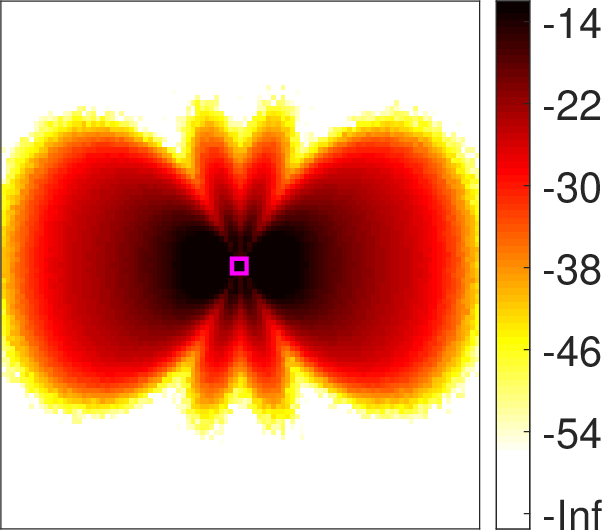}
	\caption{Heatmap of the average harvested energy in dBm under the CSI-free WET schemes: $\mathrm{APS-EMW}$ (left), $\mathrm{AA-IS}$, $\mathrm{SA}$ (middle-left), $\mathrm{AA-SS}_{\mathrm{min\ var}}$ (middle-right), $\mathrm{AA-SS}_{\mathrm{max\ E}}$ (right). } 
	\label{FigSAAA}
\end{figure}

Fig.~\ref{FigSAAA} shows the performance of above CSI-free WET schemes in terms of average harvested energy in a $10\ \mathrm{m}\times 10\ \mathrm{m}$ area. It is assumed a PB equipped with four antennas and located in the area center, while the devices' EH circuitry operates with sensitivity, saturation and conversion efficiency of $-22$ dBm, $-8$ dBm and 35\%, respectively \cite{Valenta.2014}.
Note that $\mathrm{APS-EMW}$, $\mathrm{SA}$ and $\mathrm{AA-IS}$ provide a uniform performance along the area, 
while $\mathrm{AA-SS}_\mathrm{\min var}$ and $\mathrm{AA-SS}_\mathrm{\max E}$ favor certain spatial directions. 
A quantitative information on the average energy availability in the area is shown in Fig.~\ref{FigCSF}. 
Observe that while $\mathrm{AA-SS}_\mathrm{min\ var}$ allows the EH devices to harvest  energy not less than $-24$ dBm on average in $9\%$ of the area, such coverage can be increased to $13\%$, if $\mathrm{APS-EMW}$, $\mathrm{SA}$ or $\mathrm{AA-IS}$ are used, or even to $18\%$ under $\mathrm{AA-SS}_\mathrm{max\ E}$. It is worth highlighting that $\mathrm{APS-EMW}$ is only preferable for  supporting massive EH deployments with ultra-low energy demands, while the other CSI-free schemes perform usually better under more stringent energy requirements. Also, among the discussed CSI-free schemes, $\mathrm{SA}$ is the only that requires a single RF chain for its optimum operation (full-diversity gain), thus, reducing circuit power consumption, hardware complexity and operational costs, but the corresponding gains in terms of average harvested energy are modest compared to $\mathrm{APS-EMW}$ (in the low-energy regime) and $\mathrm{AA-SS}_\mathrm{\max\ E}$.

\subsection{Positioning-based  Schemes}\label{Pcsifree}
The CSI-free WET schemes discussed above are mostly blind in the sense they exploit little or no information for performance improvements. 
Meanwhile, positioning information, which may be available in case of static or quasi-static IoT deployments, could be used to improve the WET efficiency. In fact,  $\mathrm{AA-SS}_\mathrm{min\ var}$ and  $\mathrm{AA-SS}_\mathrm{max\ E}$ can be adapted to partially take  advantage of positioning information by properly rotating the PB antenna array, and even be combined
to improve
coverage \cite{Lopez.2020}. However, such strategy is also limited to certain deployments where the devices are conveniently clustered. A more general and efficient positioning-based CSI-free scheme may be viable. Since WET-enabled setups are usually under certain LOS effect, one could exploit LOS (geometric) MIMO channels to design an efficient EB that allows powering clustered deployments as that shown in Fig.~\ref{Fig_pos}. An efficient design requires an appropriate:
\begin{figure}[t!]
	\centering
	\includegraphics[width=0.44\textwidth]{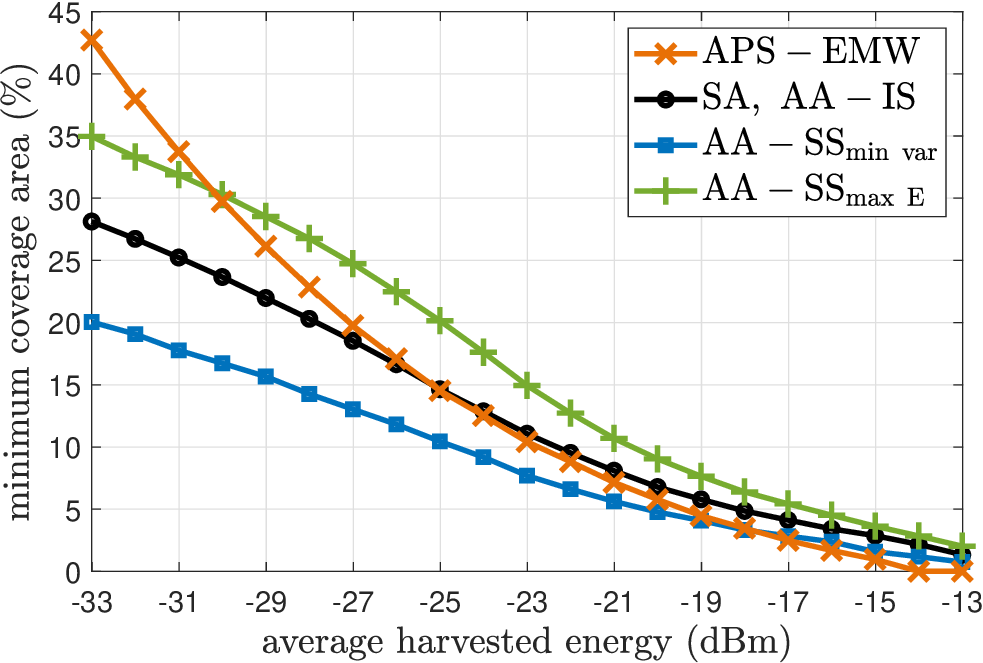}
	\caption{Area coverage for different  average harvested energy requirements.} 
	\label{FigCSF}
\end{figure}
\begin{itemize}
	\item clustering algorithm favoring the angular domain; 
	\item power control or time allocation, since clusters with the farthest devices need to be compensated with greater power/time allocation. The power control approach requires the transmission of per-cluster independent WET signals subject to a total power budget constraint to concurrently serve the entire network. Note that the maximum number of independent modulated signals that can be generated is given by the number of PB's antennas, $M$. However, such limiting phenomenon can be avoided either by transmitting non-overlapping unmodulated WET signals (deterministic multisine with different allocation of sub-carriers per cluster), or by using time allocation such that a single WET signal can be used to power each cluster at different time intervals. All these resource allocation approaches, when feasible, perform similar in terms of RF energy availability at the receiver side, but they differ in terms of EH performance when considering the non-linearities of the EH circuitry \cite{ClerckxKim.2018,Lopez.2020};		
	\item per-cluster antenna selection, since the more the number of antennas used to power certain cluster, the narrower the associated energy beam. Clusters with greater angular dispersion should be powered using wider beams, i.e., beams generated with a smaller number of antennas.
\end{itemize}
Note that positioning information is inexact in practice, and the position accuracy must be taken into account. For instance, research community and industry are targeting a positioning accuracy around 10 cm for IoT deployments in the 6G era \cite{Mahmood.2020}.
Meanwhile, although it was only emphasized  the need of a location-based clustering, efficient designs may also take advantage (when available)  of devices' battery state information, QoS requirements, e.g., to dynamically reduce the set of devices to be served and optimize the network performance. Finally, system performance is also worth  investigating when PBs are equipped with other antenna array topologies.   
\begin{figure}[t!]
	\centering
	\includegraphics[width=0.48\textwidth]{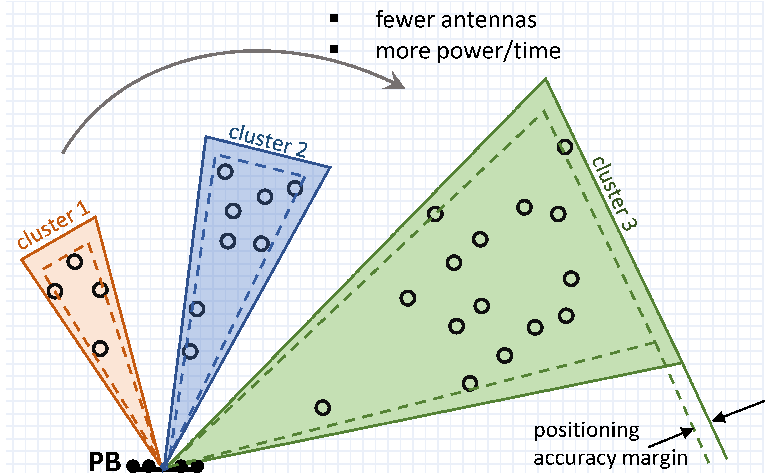}
	\caption{Example setup where the IoT deployment can be split into three clusters with 4, 8 and 15 EH devices, and the PB may use a positioning-based CSI-free EB for efficient WET. In the figure, the inner triangles (with dashed lines) delimit the region where the EH devices are predictably located, while the outer triangles (with solid lines) take into account a positioning accuracy margin. Thus, the real position of the EH devices is illustrated in the figure.}
	\label{Fig_pos}
\end{figure}
\subsection{Hybrid Schemes}\label{hybrid}
In practice, due to network heterogeneity, PB(s) might need to simultaneously power devices for which CSI acquisition procedures are, or are not, affordable, and their position information is, or is not, available. This 
may demand the adoption of proper hybrid schemes combining full-CSI, CSI-limited, CSI-free and/or positioning-based EB strategies. In this sense, appropriate designs require to deal with both the finitude of the pilot pools for CSI training  and all the available side-information. 

Meanwhile, note that CSI/positioning-based designs may be required for the PB(s) to serve intended/legitimate EH devices via ultra-narrow/precise energy beams, e.g., for dedicated energy trading. However, running CSI acquisition procedures and/or acquiring devices' positioning information, including the associated scheduling/signaling mechanisms, may be prohibited for some devices with seriously depleted batteries. In such cases, CSI-free WET schemes may be used in a first stage to lowly/controllably energize nearby EH devices. If needed, such EH devices can then establish advanced protocols, e.g., to negotiate energy provision, with the PB that allow CSI and/or positioning information acquisition, thus enabling the subsequent high-gain WET. All in all, CSI-free WET schemes may be necessary as an initial step to allow CSI/positioning-based EB in poorly energized IoT devices. How often and how long the PB should perform CSI-free WET, together with appropriate associated protocols, are open and interesting research directions.
%
%
\section{Conclusions}\label{conclusions}
WET is a promising solution for powering the  IoT in the 6G era where a huge number of devices will require steady and uninterrupted operation.
This work overviewed its distinctive features, architecture, and key IoT applications such as  RFID, live labels, wireless sensor networks and wake-up radios. Potential enablers for efficient and scalable WET were discussed. In this sense, concepts and novel ideas related to EB, DAS, devices' hardware, programmable medium, new spectrum opportunities, resource scheduling and DLT, were thoroughly revised, while highlighting the main associated challenges.
The benefits from combining DAS, via PBs deployment optimization, and EB were  illustrated via numerical results. 
Additionally,  
CSI acquisition was emphasized as a key limitation towards massive WET, thus evidencing the suitability of CSI-limited and CSI-free strategies in such scenarios.
It was shown that an EB based on average CSI can attain near optimum performance with limited overhead, hence, reduced devices' energy consumption.
Additionally, the pros and cons of the state-of-the-art CSI-free techniques, and the potential benefits from exploiting devices' positioning information, were discussed. 
As a consequence, a cluster-based EB scheme will be formally proposed in a future work, together with its integration to DAS for further improvements.

To conclude, some key challenges
and research directions identified throughout the paper are summarized in Table~\ref{challenges}.

\begin{table*}[!t]
	\centering
	\caption{Summary of some research directions}
	\vspace{-3mm}
	
	\begin{tabular}{
			L{0.19\textwidth} L{0.48\textwidth} L{0.106\textwidth}
			L{0.125\textwidth}}
		\toprule \\[-4mm]
		\textbf{Challenge} & \textbf{Candidate Approaches} & \textbf{Requirements} & \textbf{Tools}
		\\[-0.8mm] 
		\midrule	    
		\textit{Efficient EB given heterogeneous CSI availability:} In heterogeneous deployments, a PB may need to simultaneously power devices for which CSI acquisition procedures are, or are not, affordable. In the latter case, some statistical CSI may be available. & 
		Appropriate strategies are required to deal with the finitude of the pilot pools. Devices' clustering, and corresponding pilot allocation, according to network traffic features and QoS requirements may be needed. Efficient EB designs need to consider the resulting network heterogeneity.
		Soft-clustering (where there is only a probabilistic certainty of belonging to certain cluster), probabilistic optimization and novel networking mechanisms (to promote information sharing among devices) may be needed. & full/limited CSI, QoS requirements, location, EH circuit information & statistical optimization, DLT, cooperation, ML/AI, soft \& hierarchical clustering, hidden Markov models\\
		\hdashline
		\textit{Optimized DAS:} Geographically-distributed antennas improve WET efficiency by mitigating the large-scale fading and promoting LOS conditions. 
		However, efficient deployments and EB strategies are  necessary to optimize DAS performance for  low-power massive IoT.
		& 
		PBs serve EH devices clustered in their surroundings. Optimally deploying PBs requires the EH clusters to be first  efficiently identified. However, traditional clustering algorithms may be strictly sub-optimal since a PB powering certain cluster may significantly influence other nearby clusters as well. 
		Also, the available number of antennas per PB may influence the optimal clustering.
		In certain scenarios, designs may need to even account for uncertainty in devices' positioning information.
		Moreover, statistical EB must be efficiently designed to further avoid CSI acquisition procedures in DAS. Although standalone statistical beamforming designs could be adapted to DAS as centralized algorithms, the use of distributed solutions is usually preferable
		and must consider the characteristics/challenges of the specific DAS deployment, e.g., radio stripe systems.
		& 
		devices' position information and/or QoS requirements, local statistical CSI & soft \& hierarchical clustering,  meta distribution,  network calculus, optimization, distributed algorithms\\
		\hdashline	
	\textit{Efficiently adaptable WET:} 
  Efficient and scalable  WET must effortless and reliably adapt to	network dynamics and heterogeneity, e.g., 
     devices'/PBs' mobility, IRS availability, energy provision contracts. &
      Under high mobility conditions, where channel coherence time shrinks considerably and CSI gets quickly outdated, location-based EB may be mandatory.
     Meanwhile, motor/rotor-equipped PBs, UAPBs, IRS and highly directive mmWave WET, can significantly benefit from location-based EB since a strong LOS influence is usually available. More efficient CSI-free schemes and energy trading mechanisms can also be designed to exploit positioning information. Note that such location-based EB can be implemented digitally (in case of PBs equipped with multiple unidirectional antennas) and/or exploiting hardware adaptability (e.g., reconfigurable antennas or swarm-enabled virtual antennas).  In any case, an accurate prediction of the receivers' LOS direction is required, while positioning errors and/or beam-misalignment statistics must be considered. 
     & environment geometry, local/global network information, devices' location, QoS requirements & location estimation/prediction, ML/AI, optimization, game theory, ray tracing, distributed algorithms, cooperation, energy trading, DLT, MEC \\
		\hdashline				
	\textit{IRS-enabled mWET:} IRS deployments must be carefully planned, at least in a first implementation wave where IRS availability will be more limited. Novel EB designs must deal with constrained CSI availability and agree with fully passive IRS implementations.
	& Location of IRS with respect to target EH devices and PBs significantly impacts WET efficiency, thus, it needs to be properly optimized specially in static or quasi-static deployments. Risk-aware clustering and placement optimization approaches are needed. IRS' shape and number of reflective elements, together with the number of IRS that require to be deployed, need to be optimized to enable mWET. Moreover, CSI-limited EB designs may be unavoidable because of both devices' energy constraints and fully passive IRS implementations. Location-based EB designs are attractive.		
	& environment geometry, devices' location, statistical CSI, QoS requirements &  location estimation/prediction, ray tracing, materials physics,  ML/AI, statistical optimization, risk theory, soft \& hierarchical clustering \\
		\hdashline 
	\textit{Pushing the energy consumption to zero:}  
	Realizing ultra-low-power circuits/protocols is essential to fully realize the potential of WET to power massive IoT deployments. Further research on form factor miniaturization, light AI on chip, and seamless circuit integration into everyday object is needed.	
	 &
   Ultra-low-power circuits may require integrating antennas/rectennas into the device package with an optimized intimate connection, while incorporating PMUs that optimize the net harvested energy (harvested energy minus power consumption over time) to ensure real system gains. Additionally, WURs, duty cycling or event-driven architectures (with short power-up settling times to enable swift change between sleep and on states) may be adopted to reduce transceiver usage. 
   On-chip intelligence needs to be developed to run/facilitate at least small-scale tasks, e.g. smart-wake up and PMU functionalities, while large-scale tasks may be offloaded to MEC for further processing and energy saving.    
   The integration of EH circuits in  a plethora of materials, e.g., glass, textiles, skin,  and technologies such as  printing  and roll-to-roll compatible techniques, need to be further explored.
    & 	hardware/
    network/
    function context information    & MEC, low-complexity AI/ML, quantization/classification mechanisms, energy trading, risk theory, materials physics	\\	
    	\hdashline 
    \textit{Blockage and safety issues in mmWave WET:}  
   Blockage/non-LOS and occasionally severe signal attenuation conditions need to be addressed under strict radiation constraints in mmWave WET.
    &
   To overcome blockage/non-LOS conditions, or even
  occasionally severe signal attenuation, e.g., in outdoor IoT deployments under unfavorable weather conditions, multi/hybrid radio access technologies and/or DAS may need to be leveraged. Note that mmWave DAS is a promising approach to also solve safety issues as smaller path losses need to be compensated by the beam gains, and network could be even combined with sensing-based humans real-time detection technologies to cease beam-specific WET when it deems to be harmful.  
    &  devices' location, statistical CSI, QoS requirements     & multi-connectivity, location estimation/prediction, ray tracing, distributed optimization, risk theory, meta distribution	\\	
    	\hdashline 
    \textit{Sustainable WET:}  
   PBs may be powered using green sources, e.g., EH from ambient energy sources, to fully realize sustainable networks. Energy availability issues need to be seriously taken into account.
    &
    Energy resources at PBs powered  by \textit{in-situ} green energy, from either ambient or dedicated EH, are intrinsically more limited/volatile. 
    Thus, efficient distributed energy scheduling, cooperation and usage protocols may be needed to support network-wide WET from PBs to IoT devices. Also, both infrastructure and IoT nodes may trade their energy goods or surplus with other nearby nodes for which they can act as opportunistic PBs.
    DLT-based strategies may provide security and privacy in such energy transactions, which require novel EB designs to guarantee the agreed QoS levels with minimum communication overhead.
    & 	green energy sources characterization, devices' location, QoS requirements, CSI, battery evolution    & DLT, energy trading, MEC, cooperation, low-complexity optimization, game theory 	\\	
	\bottomrule		
	\end{tabular}\label{challenges}
	\vspace{-4mm}
\end{table*}

\bibliographystyle{IEEEtran}
\bibliography{IEEEabrv,references}
%
%
%
%
%
%
%
\end{document}